\RequirePackage{amsmath}
\documentclass[journal=tosc,spthm]{iacrtrans}
\usepackage{amssymb} 
\usepackage{amsfonts}
\usepackage{cite} %bibtex
\usepackage{graphicx} 
\usepackage{supertabular}

\usepackage{caption}
\usepackage{subcaption}

\usepackage{physics}
\usepackage{soul}
\usepackage{multirow}
\usepackage[export]{adjustbox}
\usepackage[english]{babel}
\usepackage{slashed}
\usepackage{amsmath}
\usepackage{mathtools}
\usepackage{qcircuit}
\usepackage{multicol}
\usepackage{listings}

\usepackage{hyperref}
\usepackage{cleveref}

\usepackage{tikz}
\usepackage{mathdots}
\usepackage{yhmath}
\usepackage{cancel}
\usepackage{color}
\usepackage{siunitx}
\usepackage{array}
\usepackage{tabularx}
\usepackage{booktabs}
\usetikzlibrary{fadings}
\usetikzlibrary{patterns}
\usetikzlibrary{shadows.blur}
\usetikzlibrary{shapes}

\usepackage{xspace}

\newcommand{\chachaperm}{\mathsf{ChaCha_{\pi}}}
\newcommand{\quarterround}{\mathsf{QR}}
\newcommand{\toysponge}{Toy Sponge Hash\xspace}
\newcommand{\toyblake}{Toy BLAKE Hash\xspace}

\begin{document}

\title{Quantum Search for Scaled Hash Function Preimages}

\authorrunning{Ramos-Calderer et al.}
%\authorrunning{Anonymous Author}
	\author{
	    %Anonymous Author
     	Sergi Ramos-Calderer\inst{1}\inst{2} \and
	    Emanuele Bellini\inst{1} \and
	    Jos\'e I. Latorre\inst{1}\inst{2}\inst{3} \and
	    Marc Manzano\inst{1} \and
	    Victor Mateu\inst{1}
	}
	\institute{
	    %Anonymous Institution
 		 Technology Innovation Institute, United Arab Emirates.  \\ \email{sergi.ramos@tii.ae, emanuele.bellini@tii.ae, jose.ignacio.latorre@tii.ae, marc@tii.ae, victor.mateu@tii.ae}
 		 \and
 		 Departament de F\'isica Qu\`antica i Astrof\'isica and Institut de Ci\`encies del Cosmos, Universitat de Barcelona, Spain.
 		 \and
 		 Centre for Quantum Technologies, National University of Singapore, Singapore.
		}

\maketitle
\begin{abstract}

We present the implementation of Grover's algorithm in a quantum simulator to perform a quantum search for preimages of two scaled hash functions, whose design only uses modular addition, word rotation, and bitwise exclusive or. Our implementation provides the means to assess with precision the scaling of the number of gates and depth of a full-fledged quantum circuit designed to find the preimages of a given hash digest.
The detailed construction of the quantum oracle shows that the presence of AND gates, OR gates, shifts of bits and the reuse of the initial state along the computation, require extra quantum resources as compared with other hash functions based on modular additions, XOR gates and rotations. We also track the entanglement entropy present in the quantum register at every step along the computation, showing that it becomes maximal at the inner core of the first action of the quantum oracle, which implies that no classical simulation based on Tensor Networks would be of relevance. Finally, we show that strategies that suggest a shortcut based on sampling the quantum register after a few steps of Grover's algorithm can only provide some marginal practical advantage in terms of error mitigation. 


\end{abstract}

\section{Introduction}

Cryptography is universally used to protect the security - confidentiality and integrity -  of communications and stored data. As it is common in the information security world, the security of a cryptographic scheme is measured by the computational cost required to recover the secret or the plaintext of the communication. 
For many years, the computational complexity was evaluated in terms of computer instructions required to run an algorithm that solves this problem. However, this paradigm has totally changed due to the fact that the technology on the quantum computers side has evolved up to a point in which they could be a reality in the next years.

The main threat that the existence of large enough quantum computers poses to cryptography is that, nowadays, 
all public key schemes that are standardized and massively used in our communications %nowadays 
will be insecure due to Shor's algorithm~\cite{Shor97}. An attacker could store the communications of today and decrypt them once he has a quantum computer with the required resources. In order to address this problem, the cryptographic community started designing quantum resistant schemes capable of sharing symmetric keys due to the robustness of these schemes against quantum attacks (i.e. so-called post-quantum cryptography).

Symmetric cryptographic primitives, such as hash functions, are believed to be quantum resistant. The security of hash functions is measured in terms of resistance against collision finding, preimage and second preimage finding, and their multi-target variants.
For an ideal cryptographic hash function providing $n$-bit security, 
the classical complexity of preimage and second preimage finding is $2^{n}$ expected oracle calls, while for collision finding is $2^{n/2}$ due to the birthday paradox. For these, the parallel rho method \cite{OorschotW94} can offer lower complexities if it is parallelized with a large amount of processors. For multi-target preimage search, the cost is $2^{n-t}$ hash outputs, where $2^t$ is the number of targets.

However, if the attacker had access to a quantum computer, the best algorithm for finding a preimage would be Grover's algorithm~\cite{search-grover1996} with complexity $2^{n/2}$ quantum evaluations. Some more specific applications of this algorithm can be used in order to find collisions~\cite{BHT16} with complexity $2^{n/3}$ quantum evaluations. Finally, for multi-target preimage the cost is $2^{(n-t)/2}$ quantum evaluations and all of them can also be parallelized~\cite{BB17}.
This improvement is relevant in terms of impact on the parameters for hash functions and symmetric encryption, but it is not as disruptive as Shor's algorithm for prime factorization and discrete logarithms.

Given that Grover's algorithm only provides at most a quadratic speed-up, the generally accepted approach 
to make symmetric ciphers or hash functions quantum resistant
is to double their classical security level. 
This only gives a rough idea of the security penalties that quantum computers cause on symmetric primitives, especially because the cost of evaluating Grover's oracle is very often ignored. 
Both cryptosystem designers and cryptanalysts may want to know the specific parameters that provide appropriate security, and to achieve that, further detailed studies are required to better understand 
the actual cost of quantum algorithms.

When speaking about complexity in the quantum setting, there are different assumptions related to the challenges of quantum technology. 
As in the classical computational model, we have a definition of complexity, with regards to the amount of operations, which helps the community in setting bounds on the computational cost of running a specific algorithm that breaks the security of a scheme. Nevertheless, the quantum setting also presents restrictions for a quantum circuit that describes an algorithm in terms of the number of qubits (i.e. width) and the time that it runs (i.e. depth), the latter as a result of the finite coherence time of physical qubits. For instance, an attacker can have a quantum computer but not with enough qubits to run an algorithm. Working with large quantum circuits presents challenges in terms of handling the complexity of the states, the demand in energy and/or qubit decoherence. Take into consideration that, as of now, the largest quantum computer is of 53 qubits~\cite{QSupremacy2019} and that this figure is increasing slowly, therefore, this measure serves as a sort of technological complexity metric.

\section{Related work}

NIST, in its Post-Quantum Cryptography (PQC) standardization process for asymmetric cryptographic primitives \cite{NIST2016}, suggests an approach where quantum attacks are restricted to a fixed running time, or quantum circuit depth, by a parameter named \texttt{MAXDEPTH} \cite{NIST2016}, which NIST considers to be the total amount of quantum computations possible during the full attack even in the case of error-free computations. Besides, this restriction might as well be motivated by the difficulty of running long serial computations on a quantum computer due to decoherence.

The restrictions from NIST motivated the need to provide better estimations of the number of quantum gates required to break either AES or SHA3. The security categories in the NIST PQC standardization process are defined based on the concrete cost of quantum resources in an exhaustive key search against AES and collision search as described in~\cite{GrasslLRS16}. In \cite{GrasslLRS16}, the authors aim at minimizing the circuit width (i.e. the number of qubits needed) when using Grover to break AES. Later on, the total number of Toffoli gates of the quantum circuit for AES-128 is reduced in \cite{AlmazrooieSAM18}. In parallel, in \cite{KimHJ18}, the time-space trade-offs for key search on block ciphers are discussed. They also consider NIST's \texttt{MAXDEPTH} and propose parallelization strategies for Grover's algorithm to address the depth constraint. Following this research line, a different S-box design is proposed in~\cite{LangenbergPS19} to reduce the total number of Toffoli gates in the S-box as well as its Toffoli depth for AES, improving the previous results of \cite{GrasslLRS16} and \cite{AlmazrooieSAM18}. In~\cite{BonnetainNS19}, a new framework mostly focused in AES to study quantum attacks is presented. The authors explore the different techniques used in classical cryptanalysis and how they can be sped up using quantum computers.

Recently, other symmetric primitives apart from AES have been considered. In~\cite{schlieper2020place} an implementation of Gimli is described with the particularity that it is conducted in-place (i.e. not using any ancillary registers), by taking a bit-by-bit approach and as a result of the underlying Gimli design which conducts XOR operations after performing ANDs and ORs. SIMON and SPECK have also been studied with the aim of providing quantum resource estimates in \cite{anand2020grover} and \cite{jang2020grover}, respectively. In addition, in~\cite{anand2020grover} a reduced version of SIMON was designed with the aim of verifying the implementation via quantum simulations.

Generic quantum cryptanalysis has also been proposed in the last years focusing on different constructions \cite{Kaplan14,KaplanLLN16, ChaillouxNS17}. Most of the works are purely theoretical due to the lack of a powerful enough quantum computer where an algorithm can be run to perform adequate measurements. Nevertheless, several simulation tools have recently been presented and could shed some light on the cost of running quantum algorithms. An implementation of the full Grover's oracle for key search on AES and LowMC in Q\# appeared in \cite{JaquesNRV20}. They offer a specific implementation that allows them to be more precise with the estimates of the resources that would be required to run the algorithm on a quantum computer. Besides, they also review the parallelization strategies to overcome the \texttt{MAXDEPTH} constraint from NIST, and conclude that it is advantageous. Finally, they propose a circuit minimizing the (a) gate-count and (b) depth-times-width cost metrics, under a depth constraint \texttt{MAXDEPTH}. It is important to note that their simulation never runs the full algorithm, but only parts of it, and thus they are capable of testing only small components of their implementation and claim its overall correctness.

\subsection{Our contribution}
\label{sec:our_contribution}
Following the idea from \cite{JaquesNRV20, anand2020grover} that experimental implementation gives a different and complementary view than a pure theoretical analysis, on quantum attacks complexity, we study the implementation of Grover's algorithm to find preimages of hash functions based on modular Addition, word Rotation, and eXclusive or (ARX) operations. 

Given the limitations to simulate large quantum computers, in this work we consider two ARX-based scaled hash functions with the objective of being able to verify our results.
The first uses a Sponge structure 
\cite{guido2011cryptographic}, 
using as permutation a scaled down version of ChaCha20 stream cipher internal permutation 
\cite{bernstein2008chacha} (for ChaCha12 permutation usages see for instance \cite{djb2015sphincs}).
The second is a scaled down version of the popular BLAKE2 hash function 
\cite{aumasson2014blake2}. 

We provide an implementation for both scaled hash functions, as well as Grover's algorithm, which allows us to provide precise quantum security bounds for the equivalent non-scaled hash functions, 
in terms of qubits and quantum gates required to find preimages. Moreover, we study the behaviour of the algorithm running on a simulated quantum computer in order to motivate different approaches in cryptanalysis using Grover's algorithm. In this context, we observe that there exists a trade-off to find preimages without having to run all the steps of Grover's algorithm, at the cost of increasing the probability of obtaining an incorrect preimage. This approach, introduced in \cite{boyer1998tight}, can also be used to obtain a solution when the number of preimages is unknown.

Furthermore, our work provides insights to better understand what type of operations require a higher amount of quantum resources. We infer that there are specific permutation constructions that have a higher impact on the complexity of the quantum circuit. 

The structure of this article reflects the methodology followed in this work. In \cref{sec:toy_hash}, we describe the scaled hash functions that we implement, and motivate the choice for such constructions. Next, in \cref{sec:attack_toy_hash} and \cref{sec:attack_toy_blake}, we explain our implementation of the quantum algorithm to break the two scaled hash functions. Finally, \cref{sec:results} and \cref{sec:conclusions} are for presenting the results of our experiments and the conclusions of this work.

\section{Toy hash functions}
\label{sec:toy_hash}

In order to have a better understanding of cryptanalytic attacks, 
it is a common practice to define scaled, or
\emph{toy}, versions of a cryptographic primitive 
preserving the same security properties, 
e.g. as in \cite{musa2003simplified}, 
and then run the attack on this toy primitive. 
This is also true for asymmetric cryptography, 
where, to estimate the actual difficulty of solving a problem, e.g. factoring, 
researchers try to solve the largest possible instances of that problem and 
accordingly extrapolate to measure the actual hardness of the larger problem.

In this work, we 
implement Grover's search algorithm on a simulator of a quantum computer, 
in order to understand what are the challenges of such an implementation. 
In particular, we use Grover's algorithm 
to find all possible preimages 
(with size up to the input block) 
of a hash function. 
Due to the currently limited number of qubits that can be simulated on classic computers, 
it is particularly hard to implement 
any real cryptographic primitive 
using a quantum computer simulator. 
Therefore, we consider a scaled version of a hash function, preserving the original design choices.
Among all possible hash designs available, 
we select the ones that seem easier 
to scale and fit in the simulator. 
In particular, we try to avoid AND gates and OR gates as they require additional qubits and can easily increase the qubit count over the maximum threshold that the simulator can deal with 
(see also \cref{sec:toysponge_circuit}).
Therefore, we focus on primitives using only 
modular Addition, 
bitwise Rotation and 
bitwise eXclusive or operations. 
This type of primitives are usually referred to as ARX.

In \cref{sec:toy_sponge}, 
we describe a toy hash function, referred to as \textit{\toysponge} in this article, based on the Sponge construction with an iterated
permutation
\cite[Chapter 8]{guido2011cryptographic}.
In \cref{sec:toy_blake2}, 
we describe a toy version of the popular BLAKE2 hash function
\cite{aumasson2014blake2,blake2rfc}, which we refer to as \textit{\toyblake}.

\subsection{\toysponge}
\label{sec:toy_sponge}

\begin{figure}[ht]
\centering     
\begin{subfigure}[b]{0.71\textwidth}
    \includegraphics[width=\textwidth]{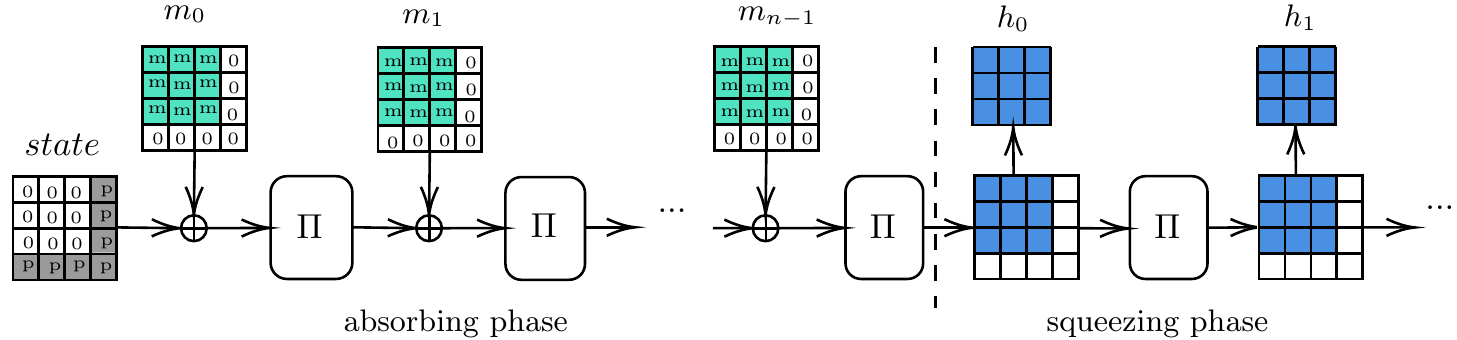}
    \caption{Full Sponge Hash (processing $n$ message blocks).}
    \label{fig:sponge_hash}
\end{subfigure}
\hfill
\begin{subfigure}[b]{0.23\textwidth}
    \includegraphics[width=\textwidth]{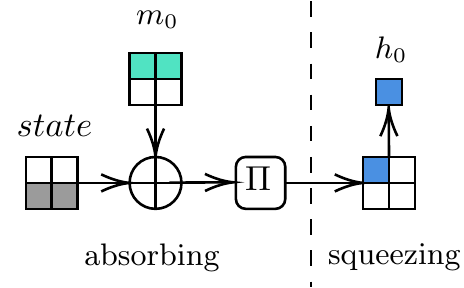}
    \caption{\toysponge (processing $1$ message block).}
    \label{fig:toy_sponge_hash}
\end{subfigure}
\caption{Sponge hash constructions based on the permutation $\Pi$.}
\end{figure}

A Sponge function \cite{guido2011cryptographic} 
is a very flexible and elegant cryptographic design, from which it is possible to derive several cryptographic primitives, including hash functions.
The security of a Sponge function is based on the security of an internal function, which is often selected as an iterated permutation $\Pi$.
In \cref{fig:sponge_hash} 
we show the diagram of a Sponge-based cryptographic hash.
During the \emph{absorbing phase}, 
the message is split in $n$ blocks $m_0,\dots,m_{n-1}$ of size $r$, 
the \emph{rate} of the Sponge, and 
xored with the Sponge state of size $r+c$, where 
$c$ is called the \emph{capacity} of the Sponge.
Note that $r+c$ is also the input/output size of the permutation $\Pi$.
The state is initialized to a fixed public value for the first iteration.
Once the message blocks are all injected into the Sponge, the \emph{squeezing phase} starts, 
producing as many output blocks, $h_0,h_1,\ldots$ of size $r$, 
as needed to reach the hash digest size.

As a first experiment, we 
implement a simple ARX permutation, 
whose design is derived from the internal permutation used in ChaCha20 stream cipher
\cite{bernstein2008chacha}. 
The state of ChaCha20 internal permutation is seen as a 4x4 matrix of 32 bits words. 
This permutation is obtained by iterating a certain number of rounds.
Each round alternatively acts on the columns and on the diagonals of the state, 
by applying a function usually referred to as the \emph{Quarter Round} ($\quarterround$).
Since our quantum simulator can only manage a 16 bit state, we define \toysponge, whose state is a 2x2 matrix of 4 bit words, 
and define the $\quarterround$ function to accept two input words and return two output words (instead of four), as shown below:
% \nolinenumbers

% \noindent
% {\scriptsize
% \begin{minipage}[t]{.32\textwidth}
% \begin{lstlisting}[escapeinside={(*}{*)}]
% (*$\quarterround$*)(a,b):
%   a = a + b
%   b = (b ^ a) <<< 2 
%   a = a + b
%   b = (b ^ a) <<< 1
%   return a, b
% \end{lstlisting}
% \end{minipage}
% \begin{minipage}[t]{.32\textwidth}
% \begin{lstlisting}[escapeinside={(*}{*)}]
% ColQR(v):
%   # update columns
%   v[0],v[2] = (*$\quarterround$*)(v[0],v[2])
%   v[1],v[3] = (*$\quarterround$*)(v[1],v[3])
%   return v
% \end{lstlisting}
% \end{minipage}
% \begin{minipage}[t]{.32\textwidth}
% \begin{lstlisting}[escapeinside={(*}{*)}]
% DiagQR(v):
%   # update diagonals
%   v[0],v[3] = (*$\quarterround$*)(v[0],v[3])
%   v[1],v[2] = (*$\quarterround$*)(v[1],v[2])
%   return v
% \end{lstlisting}
% \end{minipage}
% }
% \linenumbers

\noindent
{\scriptsize
\begin{minipage}[t]{.32\textwidth}
\begin{flushleft}
$\begin{aligned}
& \quarterround(a,b): \\
& \quad  a = a + b \\
& \quad  b = (b \oplus a) \lll 2  \\
& \quad  a = a + b \\
& \quad  b = (b \oplus a) \lll 1 \\
& \quad  \textbf{return } a, b \\
\end{aligned}$
\end{flushleft}
\end{minipage}
\begin{minipage}[t]{.32\textwidth}
\begin{flushleft}
$\begin{aligned}
& \mathsf{ColQR}(v): \\
& \quad  \text{\# update columns} \\
& \quad  v[0],v[2] = \quarterround(v[0],v[2]) \\
& \quad  v[1],v[3] = \quarterround(v[1],v[3]) \\
& \quad  \textbf{return } v \\
\end{aligned}$
\end{flushleft}
\end{minipage}
\begin{minipage}[t]{.32\textwidth}
\begin{flushleft}
$\begin{aligned}
& \mathsf{DiagQR}(v): \\
& \quad  \text{\# update diagonals} \\
& \quad  v[0],v[3] = \quarterround(v[0],v[3]) \\
& \quad  v[1],v[2] = \quarterround(v[1],v[2]) \\
& \quad  \textbf{return } v \\
\end{aligned}$
\end{flushleft}
\end{minipage}
}

\noindent
where $\mathsf{ColQR}$ and $\mathsf{DiagQR}$ describe how $\quarterround$ is applied to the columns and diagonals of the state, respectively.

To simplify the quantum simulation,
we assume that 
the absorption of an 8 bit message 
and the squeezing of a 4 to 8 bit hash output
is performed in a single iteration of the internal permutation, as shown in \cref{fig:toy_sponge_hash}.
This means 
the rate $r$ of \toysponge is 8 bits and 
the capacity $c$ is also 8 bits.
We will use $\chachaperm$ to indicate the permutation used in \toysponge. 

% \begin{figure}
% \include{figures/sponge_with_2x2_block_one_iteration}
% \caption{\toysponge construction based on the permutation $\Pi$ (processing $1$ message block).}
% \label{fig:toy_sponge_hash}
% \end{figure}

\subsection{\toyblake}
\label{sec:toy_blake2}

\begin{figure}[ht]
\centering   
\begin{subfigure}[b]{0.49\textwidth}
    \includegraphics[width=\textwidth]{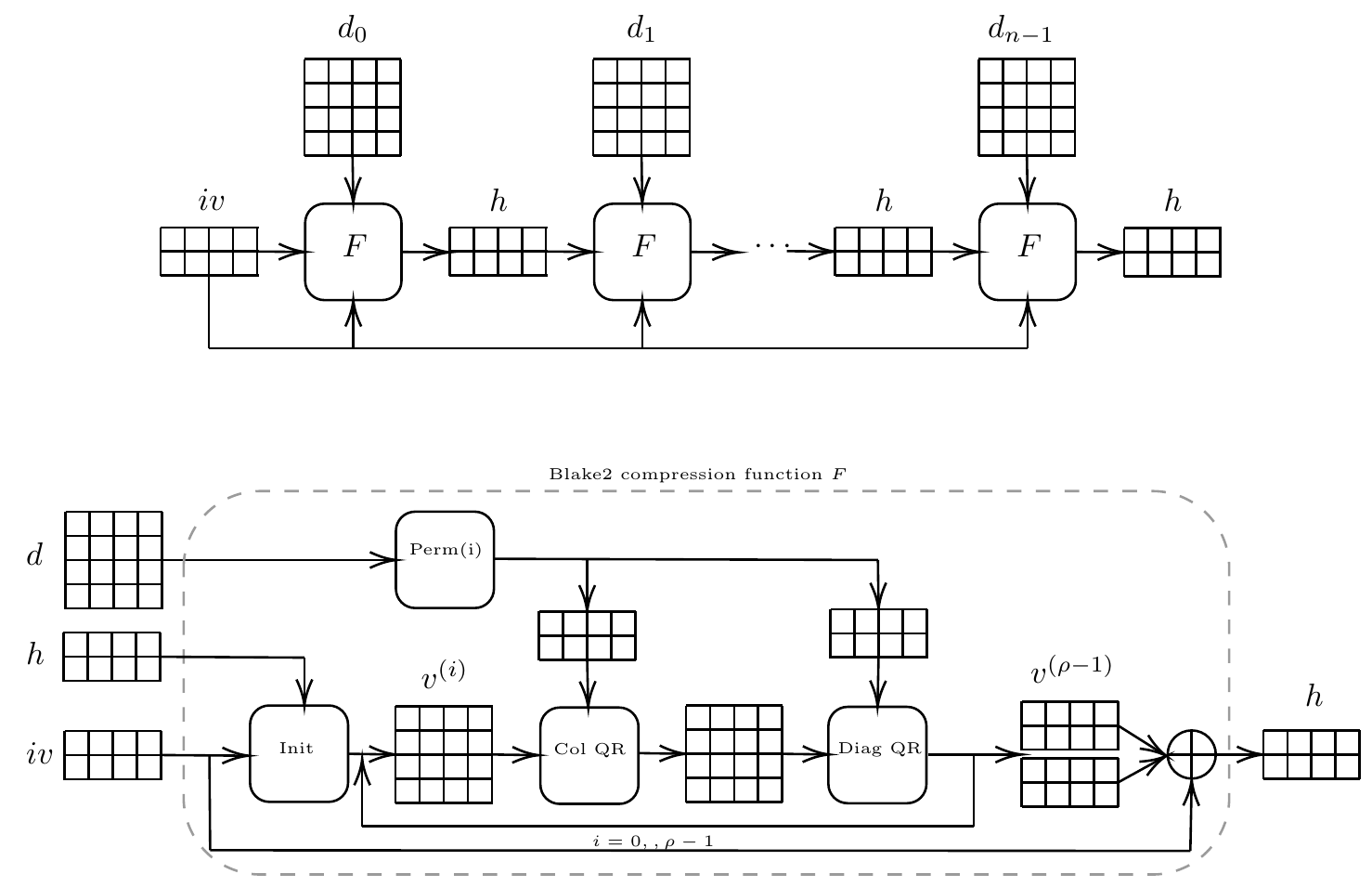}
    \caption{BLAKE2 hash function and its internal compression function $F$.}
    \label{fig:blake2_compression_function}
\end{subfigure}
\hfill
\begin{subfigure}[b]{0.49\textwidth}
    \includegraphics[width=\textwidth]{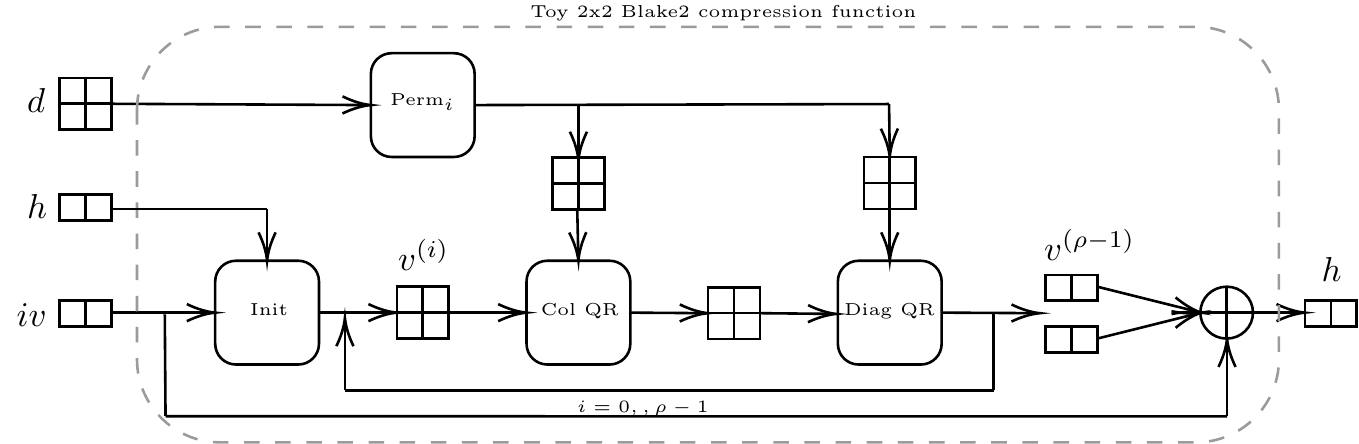}
    \caption{\toyblake compression function, with a 2x2 matrix state.}
    \label{fig:blake2_toy2x2_compression_function}
\end{subfigure}
\caption{BLAKE2 structure.}
\end{figure}

BLAKE2 \cite{aumasson2014blake2,blake2rfc} 
is an ARX cryptographic hash function designed to have the best performances on software implementations. 
Its core permutation is a tweak of ChaCha20 stream cipher permutation
\cite{bernstein2008chacha}. 
To produce a message digest, 
a message is split in $n$ blocks $d_0,\ldots, d_{n-1}$, 
consisting of 4x4 matrices of 64 (or 32) bit words.
Each block is input into a compression function $F$ together with an 8 words state $h$, 
which is then updated by $F$.
The compression function $F$ initializes a 4x4 matrix $v$ 
using $h$ and a public initialization vector $iv$. 
Then, for $\rho$ rounds, 
$F$ acts first on the columns and then on the diagonals of $v$, 
by applying a mixing function $G$, 
which also accepts a permutation of the current block as input.
The hash function BLAKE and its compression function $F$ 
are depicted in \cref{fig:blake2_compression_function}.

% PACKAGE SUBFIGURE
% \begin{figure}[ht]
% \centering     
% \subfigure[BLAKE2 hash function and its internal compression function $F$.]{\label{fig:blake2_compression_function}
% \includegraphics[width=.48\textwidth]{figures/blake2_and_compression_function.pdf}}
% % \hspace{.03\textwidth}
% \subfigure[\toyblake compression function, with a 2x2 matrix state.]{\label{fig:blake2_toy2x2_compression_function}
% \includegraphics[width=.48\textwidth]{figures/blake2_toy2x2_compression_function.pdf}}
% \caption{BLAKE2 structure.}
% \end{figure}

% \begin{figure}
% \include{figures/blake2_toy2x2_compression_function}
% \caption{\toyblake compression function, with a 2x2 matrix state.}
% \label{fig:blake2_toy2x2_compression_function}
% \end{figure}

% \begin{figure}
% \include{figures/blake2_and_compression_function}
% \caption{BLAKE2 hash function and its internal compression function $F$.}
% \label{fig:blake2_compression_function}
% \end{figure}

To derive \toyblake, the scaled version of BLAKE2, we reduce the word size to 4 bits, and the message blocks are represented as 2x2 matrices with 4 bit words entries, 
as shown in \cref{fig:blake2_toy2x2_compression_function}.
The values $h$, $iv$, and $v$ are initialized as follows:
% \noindent
% {\scriptsize
% \begin{minipage}[t]{.45\textwidth}
% \begin{lstlisting}
% iv[0] = 0x8
% iv[1] = 0xB
% h[0] = (iv[0] ^ 0x2)
% h[1] = iv[1]
% \end{lstlisting}
% \end{minipage}
% % and then, every time a round begins $v$ is initialized as:
% \begin{minipage}[t]{.45\textwidth}
% \begin{lstlisting}
% v[0] = h[0] ^ t
% v[1] = h[1] ^ (t >> 4)
% v[2] = iv[0]
% v[3] = iv[1]
% if last round then 
%   v[2] = v[2] ^ 0xF # Invert all bits.
% \end{lstlisting}
% \end{minipage}
% }

% \noindent
% {\scriptsize
% \begin{minipage}[t]{.32\textwidth}
% \begin{flushleft}
% $\begin{aligned}
% iv[0] & = \texttt{ 0x8} \\
% iv[1] & = \texttt{ 0xB} \\
% h[0] & = (iv[0] \oplus \texttt{ 0x2}) \\
% h[1] & = iv[1] \\
% \end{aligned}$
% \end{flushleft}
% \end{minipage}
% % and then, every time a round begins $v$ is initialized as:
% \begin{minipage}[t]{.32\textwidth}
% \begin{flushleft}
% $\begin{aligned}
% v[0] & = h[0] \oplus \\
% v[1] & = h[1] \oplus (t \gg 4) \\
% v[2] & = iv[0] \\
% v[3] & = iv[1] \\
% \end{aligned}$
% \end{flushleft}
% \end{minipage}
% \begin{minipage}[t]{.32\textwidth}
% \begin{flushleft}
% $\begin{aligned}
% & \textbf{if } \text{last round } \textbf{then}  \\
% & \quad  v[2] = v[2] \oplus 
% \texttt{ 0xF } \text{\# Invert all bits}. \\
% \end{aligned}$
% \end{flushleft}
% \end{minipage}
% }
\noindent
$
iv[0] = \texttt{ 0x8},
iv[1] = \texttt{ 0xB},
h[0] = (iv[0] \oplus \texttt{ 0x2}),
h[1] = iv[1],
v[0] = h[0] \oplus,
v[1] = h[1] \oplus (t \gg 4),
v[2] = iv[0],
v[3] = iv[1]
$
and for the last round all bits are inverted, i.e.
$v[2] = v[2]$.
The variable $t$ identifies the current 2x2 block that is being processed, and $v$ is initialized every time a round begins.

% \begin{figure}
% \include{figures/blake2_toy2x2_compression_function}
% \caption{\toyblake compression function, with a 2x2 matrix state.}
% \label{fig:blake2_toy2x2_compression_function}
% \end{figure}

Finally, the internal mixing function $G$ of the \toyblake quarter round is defined as follows:

% {\scriptsize
% \noindent
% \begin{minipage}[t]{.20\textwidth}
% \begin{lstlisting}
% G(a,b,x,y):
%   a = a + b + x 
%   b = (b ^ a) >>> 2 
%   a = a + b + y
%   b = (b ^ a) >>> 1
%   return a, b
% \end{lstlisting}
% \end{minipage}
% \begin{minipage}[t]{.39\textwidth}
% \begin{lstlisting}
% ColQR(v,d):
%   # update columns
%   v[0],v[2] = G(v[0],v[2], d[s[0]], d[s[1]])
%   v[1],v[3] = G(v[1],v[3], d[s[2]], d[s[3]])
%   return v
% \end{lstlisting}
% \end{minipage}
% \begin{minipage}[t]{.39\textwidth}
% \begin{lstlisting}
% DiagQR(v,d):
%   # update diagonals
%   v[0],v[3] = G(v[0],v[3], d[s[0]], d[s[1]])
%   v[1],v[2] = G(v[1],v[2], d[s[2]], d[s[3]])
%   return v
% \end{lstlisting}
% \end{minipage}
% }
{\scriptsize
\noindent
\begin{minipage}[t]{.20\textwidth}
\begin{flushleft}
$\begin{aligned}
& G(a,b,x,y): \\
& \quad a = a + b + x \\
& \quad b = (b \oplus a) \ggg 2 \\
& \quad a = a + b + y \\
& \quad b = (b \oplus a) \ggg 1 \\
& \quad \textbf{return } a, b \\
\end{aligned}$
\end{flushleft}
\end{minipage}
\begin{minipage}[t]{.39\textwidth}
\begin{flushleft}
$\begin{aligned}
& \mathsf{ColQR}(v,d): \\
& \quad   \text{\# update columns} \\
& \quad   v[0],v[2] = G(v[0],v[2], d[s[0]], d[s[1]]) \\
& \quad   v[1],v[3] = G(v[1],v[3], d[s[2]], d[s[3]]) \\
& \quad   \textbf{return } v \\
\end{aligned}$
\end{flushleft}
\end{minipage}
\begin{minipage}[t]{.39\textwidth}
\begin{flushleft}
$\begin{aligned}
& \mathsf{DiagQR}(v,d): \\
& \quad   \text{\# update diagonals} \\
& \quad   v[0],v[3] = G(v[0],v[3], d[s[0]], d[s[1]]) \\
& \quad   v[1],v[2] = G(v[1],v[2], d[s[2]], d[s[3]]) \\
& \quad   \textbf{return } v \\
\end{aligned}$
\end{flushleft}
\end{minipage}
}

\noindent
where $\mathsf{ColQR}$ and $\mathsf{DiagQR}$ describe how $G$ is applied to the columns and diagonals of the state, respectively.

\section{Quantum attack on \toysponge}
\label{sec:attack_toy_hash}

%Quantum overview of the algorithm
In order to design a quantum attack on the \toysponge construction described in \cref{sec:toy_sponge}, we need to construct an explicit quantum circuit that implements Grover's search algorithm \cite{search-grover1996}. This requires coding an oracle that performs the $\chachaperm$ as well as the different sponge phases using quantum gates, which are reversible by nature. Our quantum algorithm is developed using the Qibo quantum simulation language, available in \cite{qibo20}, and code to reproduce the examples presented in what follows can be found on Github \cite{hash-code}.
%will be uploaded to GitHub\footnote{The reference to the GitHub repository is currently removed for anonymity.}.

A real Sponge-based hash function might require several rounds of squeezing to output the digest with the expected length. In such a case, the preimage would be obtained as a backwards sequence of quantum attacks, finding preimages of preimages. Here we concentrate on the quantum algorithm for a single step in such a strategy. Solving a Sponge-based hash function in one shot would require quantum resources to store the full message and depth to accommodate all steps at once, therefore not feasible in the near term.

Next, we present the overall structure of the quantum circuit designed to attack \toysponge, and then proceed to detail its sub-parts.

\subsection{Grover's algorithm for finding preimages}

%Overview of Grover's algorithm
Grover's search algorithm \cite{search-grover1996} can be adapted to find the preimage of a hash with $2^{n/2}$ evaluations of a quantum oracle plus a diffusion operator. The specific hash function chosen must be coded onto the quantum oracle.  

The key idea behind the quantum advantage of this approach is that a quantum oracle can process calls of superposed states, hence exploiting the genuine quantum properties of entanglement and interference. It may be argued that quantum mechanics allows to try all possible preimages in parallel at a time, but needs a way to single out the desired solution. This task is non-trivial as the description of the states remains probabilistic.
A high-level understanding of the workings of Grover's algorithm comes down to the appreciation that probability amplitudes for each possible solution can add and subtract (in general, with arbitrary relative phases). It is the fact that probability amplitudes can cancel that allows for
the suppression of undesired solutions while the probability of success is enhanced beyond classical means.

Let us be more precise and specify the principal elements in Grover's algorithm, namely the initialization, the {\sl oracle} and the {\sl diffusion} operator.  

We first need to initialize the quantum register 
with a quantum superposition of all possible states. 
This is a standard step for many quantum algorithms which is achieved by applying a Hadamard gate for each qubit in the quantum register. Note that this first step is genuinely quantum, as the register will then handle the equal superposition of all possible states.

We then apply an oracle that encodes the action of the hash function. This oracle changes the sign of the states that satisfy a given condition. 
We here choose to change the sign of the hash preimage we want to unveil. Therefore, the oracle will receive all possible preimages on superposition, will compute their hash on a single go and then detect the one we want to invert. It will then be possible to change the sign of the correct preimage in the superposition and undo the hash by applying the circuit in reverse.
This means the oracle will include the information of the particular
output we are analyzing. 

After the action of the oracle, a diffusion operator is applied. This final element is constructed so as to produce an inversion of all probability amplitudes with respect to their average. The effect of the oracle plus diffusion amplifies the probability of measuring one of the solution states by a small quantity. The oracle and diffusion steps must be iterated to bring the probability of finding the right preimage close to 1.

\begin{figure}
\centering
\resizebox{0.5\textwidth}{!}{
\input{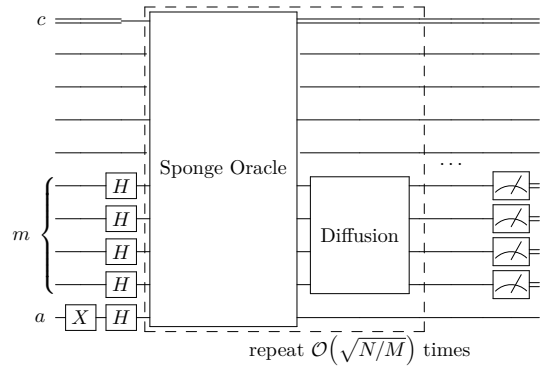}}
\caption{Scheme of a quantum circuit that would find preimages of a known hash function. The oracle must encode the \toysponge operations and flip the sign of the preimages which are searched. Each quantum wire represented in the figure corresponds to two qubits in the \toysponge implementation. The qubit register $m$ encodes the message, $c$ is a classical register and $a$ marks the Grover ancilla. The diffusion part produces the inversion over the average. The dashed box of the circuit has to be repeated $\order{\sqrt{N/M}}$ times, where $N=2^n$ is the full message space and $M$ is the number of preimages, in order to find a preimage with probability close to 1.}
\label{fig:grover_compact}
\end{figure}

% \begin{figure}
% \centering
% \resizebox{0.3\textwidth}{!}{
% \input{figures/grover_compact}}
% \caption{Scheme of a quantum circuit that would find preimages of a known hash function. The oracle must encode the \toysponge operations and flip the sign of the preimages which are searched. Each quantum wire represented in the figure corresponds to two qubits in the \toysponge implementation. The qubit register $m$ encodes the message, $c$ is a classical register and $a$ marks the Grover ancilla. The diffusion part produces the inversion over the average. The dashed box of the circuit has to be repeated $\order{\sqrt{N/M}}$ times, where $N=2^n$ is the full message space and $M$ is the number of preimages, in order to find a preimage with probability close to 1.}
% \label{fig:grover_compact}
% \end{figure}

We reiterate here the logic of the Grover attack on a hash function. Starting from equal probability amplitudes for all states, the action of an oracle, that we shall label here as "Sponge Oracle", inverts the sign for the preimage solution to \toysponge. This still remains a small probability amplitude. In the diffusion step, the inverse about the average produces an amplification of the probability amplitude associated to the solutions we are after. If a series of two-step oracle plus diffusion actions are applied $\order{2^{n/2}}$ times, a solution of the problem is found with near 1 probability.

The approach we have just sketched can be applied to search for preimages of a hash function. A diagram of the structure of Grover's algorithm to solve the \toysponge model is shown in Figure \ref{fig:grover_compact}. 

The simplicity of Grover's overall structure disappears when this algorithm is translated into a series of quantum gates to be run on a real device. Hence, the explicit coding of Grover's recipe needs to be done efficiently in order not to hinder its promised quadratic performance improvement. A detailed discussion of the creation of the oracle to solve this toy model will now follow. 

\subsection{Quantum circuit for \toysponge permutation}
\label{sec:toysponge_circuit}

\begin{figure}
\centering
\begin{subfigure}[t]{.49\textwidth}
    \includegraphics[width=\linewidth]{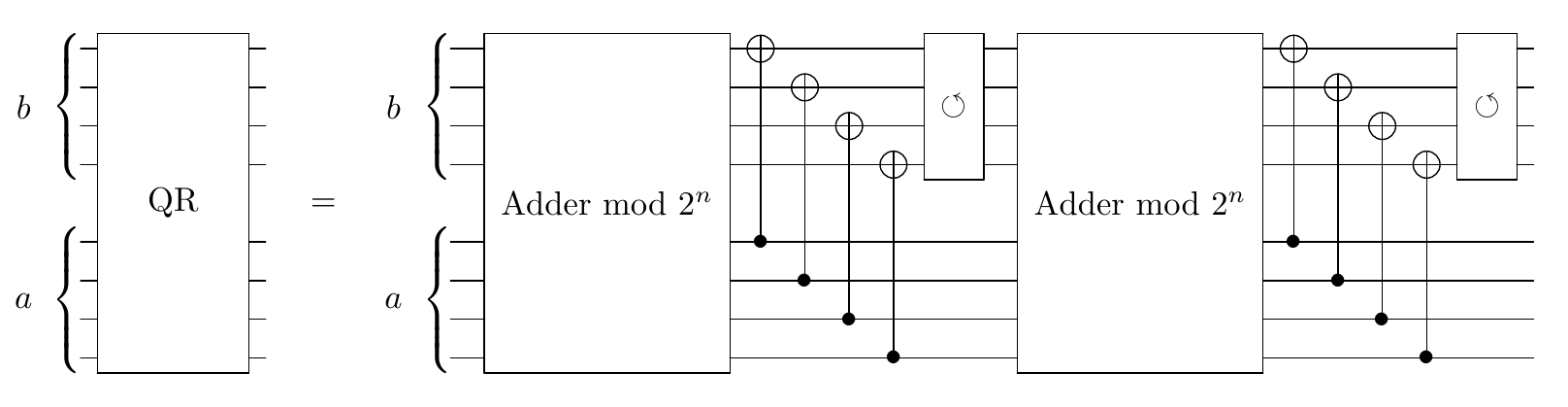}
    \caption{Quantum circuit that computes one step in the \toysponge permutation Quarter Round between registers $a$ and $b$.}
    \label{fig:qcircuit-qr}
\end{subfigure}
\hfill
\begin{subfigure}[t]{.49\textwidth}
    \includegraphics[width=\linewidth]{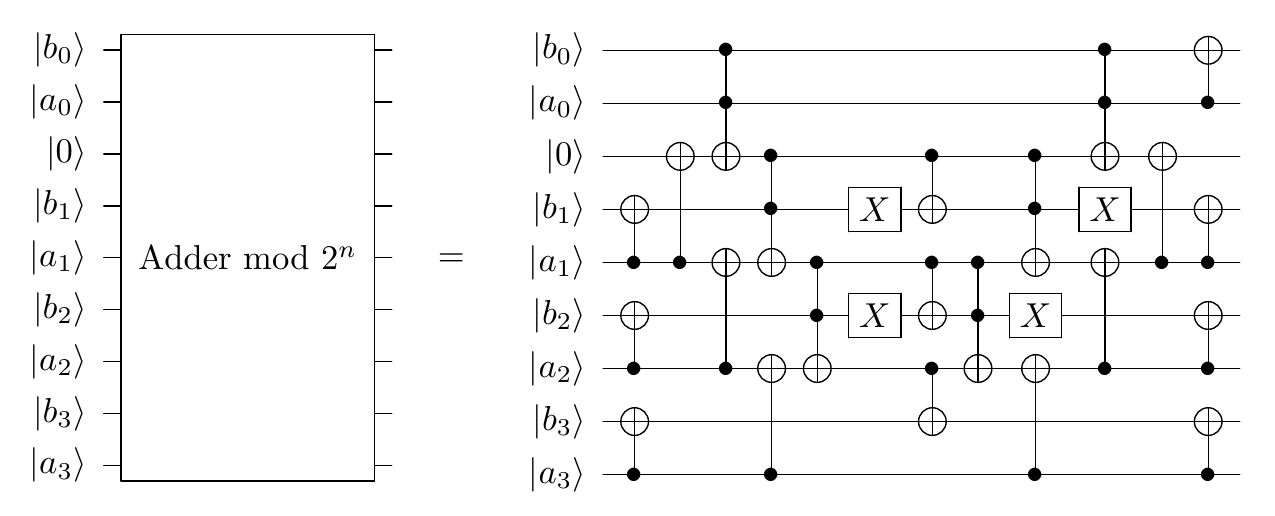}
    \caption{Quantum circuit example for the adder modulo $2^n$ between two 4 qubit register using a single ancilla qubit. This is the circuit that performs the addition part of the quantum implementation of $\quarterround$.}
    \label{fig:adder}
\end{subfigure}
\caption{Quantum circuits for the Quarter Round and the adder module $2^n$.}
\end{figure}

%Quantum implementation of the ChaCha permutation on a circuit
The base of the $\chachaperm$ permutation is the $\quarterround$ module outlined in \cref{sec:toy_sponge}. The first step in building the needed quantum Sponge Oracle is to reproduce the Quarter Round algorithm on a quantum computer. As sketched in \cref{fig:qcircuit-qr} there are three operations in $\quarterround$: an addition modulo $2^n$, a bitwise XOR operation and an $n$-bit word rotation. In terms of explicit quantum operations, the addition is the one that incurs most of the computational cost. A bitwise XOR can be achieved in a reversible manner using controlled-not (CNOT) gates and the rotation can be understood as a classical qubit relabelling and does not add any quantum cost. 

% \begin{figure}
% \centering
% \resizebox{0.5\textwidth}{!}{
% \input{figures/quarter_rotation_chacha}}
% \caption{Quantum circuit that computes one step in the \toysponge permutation Quarter Round between registers $a$ and $b$.}
% \label{fig:qcircuit-qr}
% \end{figure}

The explicit circuit design of ARX based hash functions highlights some of the issues one might face when translating a classical permutation into a reversible quantum language. As previously stated, XOR operations can be substituted by a CNOT gate, reversible due to its quantum nature. That, however is not the case for the AND and OR classical gates, as they require additional quantum resources to be added into the circuit in order to be reversible. A similar thing happens with bit shifts. While rotations can be substituted by qubit relabeling, shifts are innately destructive, therefore non-reversible, and could also need the addition of auxiliary quantum registers. The modular addition might be costly, but the ancilla qubits required to keep track of the carry bits can ultimately be decoupled from the system. 

The Quarter Round circuit we have designed makes use of an addition modulo $2^n$. This element can be constructed using a regular quantum adder without computing overflow qubits. Quantum adders have been previously studied and different algorithms are available \cite{qarithmetic-vedral1996, adder-cuccaro2004, draper2004logadder} with different depth and qubit requirements. For the purpose of this simulation we have used a modified version of the adder presented in \cite{adder-cuccaro2004} due to the reduced amount of ancillas required. As seen in \cref{fig:adder}, the circuit is highly parallelizable, enabling reduction of circuit depth and requiring one ancillary qubit.

% \begin{figure}
% \centering
% \resizebox{0.5\textwidth}{!}{
% \input{figures/adder_mod2n}}
% \caption{Quantum circuit example for the adder modulo $2^n$ between two 4 qubit register using a single ancilla qubit. This is the circuit that performs the addition part of the quantum implementation of $\quarterround$.}
% \label{fig:adder}
% \end{figure}

For each addition performed in parallel one extra ancillary qubit is needed. However, as the ancilla is decoupled from the system by the end of the computation, that same ancilla can be reused throughout the full circuit.

\begin{figure}
\centering
\begin{subfigure}[b]{0.49\textwidth}
    \centering
    \includegraphics[width=\textwidth]{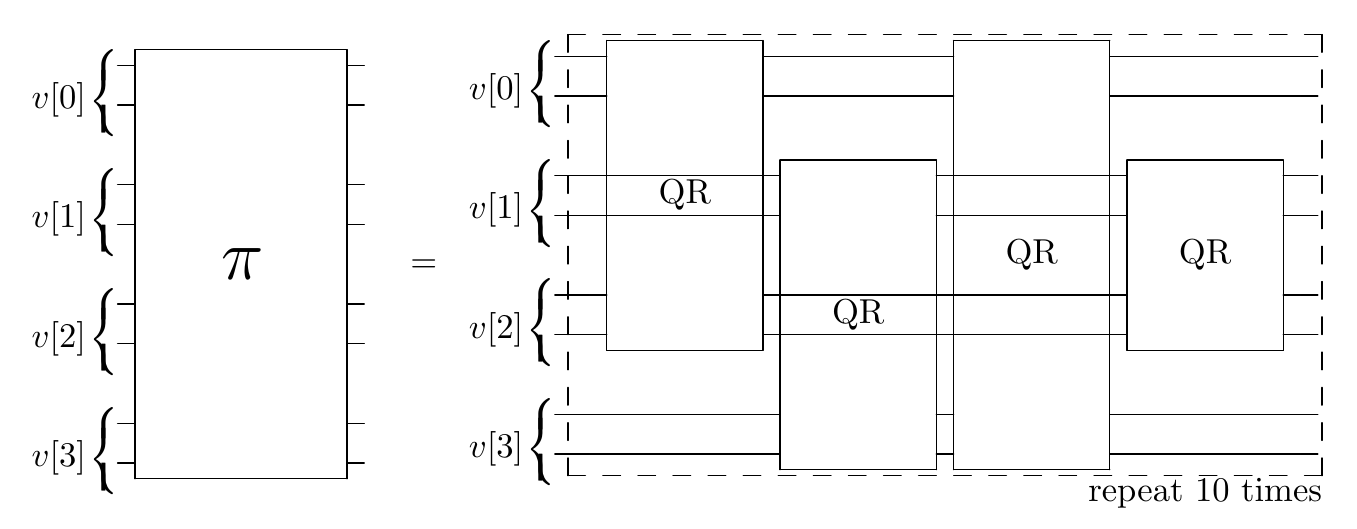}
    \caption{Quantum circuit that recreates the $\chachaperm$ permutation described in \cref{sec:toy_sponge}.}
    \label{fig:chacha-circ}
\end{subfigure}
\hfill
\begin{subfigure}[b]{0.49\textwidth}
    \centering
    \includegraphics[width=\textwidth]{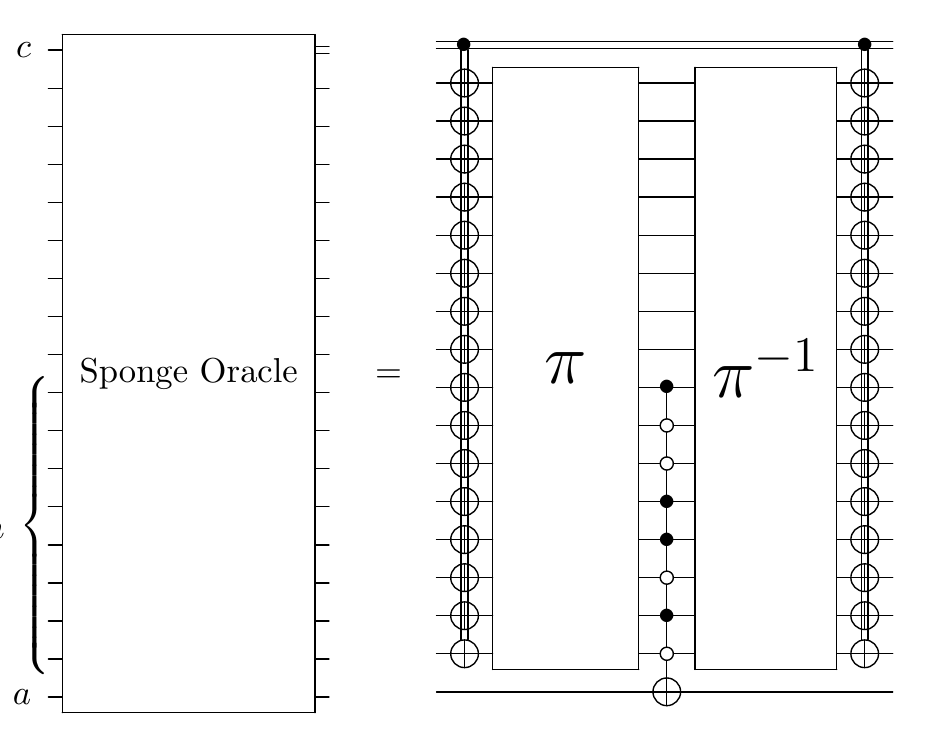}
    \caption{Sponge Oracle in the context of the \toysponge model described in \cref{sec:toy_sponge}.}
    \label{fig:oracle}
\end{subfigure}
\caption{Quantum circuit for $\chachaperm$ and for the Sponge Oracle}
\end{figure}

% \begin{figure}
% \centering
% \resizebox{0.75\textwidth}{!}{
% \input{figures/ChaCha_circuit}}
% \caption{Quantum circuit that recreates the $\chachaperm$ permutation described in \cref{sec:toy_sponge}. The Quarter Round circuits, as built in \cref{fig:qcircuit-qr}, are applied as dictate $ColQR$ and $DiagQR$ then repeated 10 times. In this figure, each visible quantum wire accounts for 2 quantum wires in the \toysponge construction and are reduced for visual clarity.}
% \label{fig:chacha-circ}
% \end{figure}

The described quantum Quarter Round block can then be added between the different quantum registers as instructed by $ColQR$ and $DiagQR$ in order to build an operator $\pi$ that outputs a $\chachaperm$ permutation on the quantum registers. The construction of this circuit is showcased in \cref{fig:chacha-circ}, where the explicit distribution of Quarter Rounds can be seen. 
The Quarter Round circuits, as built in \cref{fig:qcircuit-qr}, are applied as dictate $\mathsf{ColQR}$ and $\mathsf{DiagQR}$, 
then repeated 10 times. In this figure, each visible quantum wire accounts for 2 quantum wires in the \toysponge construction and are reduced for visual clarity.
Since the construction has been done using quantum gates, the operator will be reversible. That is, applying the gates in reverse order will recover the inverse permutation. 

\subsubsection{Full oracle.}

The full oracle we need to implement consists of three parts. 1) The first permutation, constructed classically as it does not include the message, has to be XOR-ed to the messages in superposition. 2) Then the previously described permutation is applied, and a multi-CNOT gate with controls matching the desired hash value acts on the output of the permutation and the ancilla. This step changes the sign of the quantum state that encodes the desired hash. 3) After that, the permutation is inverted in order to return to the original message space. Note that applying the permutation circuit in reverse order achieves the inverse permutation. At the end of the oracle action, all messages that output the same hash value have their amplitude sign inverted. Shown in \cref{fig:oracle} is the construction of the full Sponge Oracle using the previously described circuits.
This explicit circuit construction inverts the sign of all messages that output the same hash function in the context of the \toysponge model described in \cref{sec:toy_sponge}.
The $\pi$ operator is the quantum version of the $\chachaperm$ permutation. The $c$ wire denotes a static classical channel that determines the position of some gates, $m$ refers to the qubit register that encodes the message and $a$ labels the Grover ancilla. The hash value checked in this particular example would be $10011010$, this is determined by the controls in the multi-controlled NOT gate in the center.

%Grover oracle and its parts
% \begin{figure}
% \centering
% \resizebox{0.5\textwidth}{!}{
% \input{figures/oracle}}
% \caption{Explicit circuit construction that inverts the sign of all messages that output the same hash function in the context of the \toysponge model described in \cref{sec:toy_sponge}. The $\pi$ operator is the quantum version of the $\chachaperm$ permutation. The $c$ wire denotes a static classical channel that determines the position of some gates, $m$ refers to the qubit register that encodes the message and $a$ labels the Grover ancilla. The hash value checked in this particular example would be $10011010$, this is determined by the controls in the multi-controlled NOT gate in the center.}
% \label{fig:oracle}
% \end{figure}

\subsubsection{Diffusion operator.}
%Further explanation of the Diffusion operator

\begin{figure}
    \centering
    \includegraphics[width=0.5\textwidth]{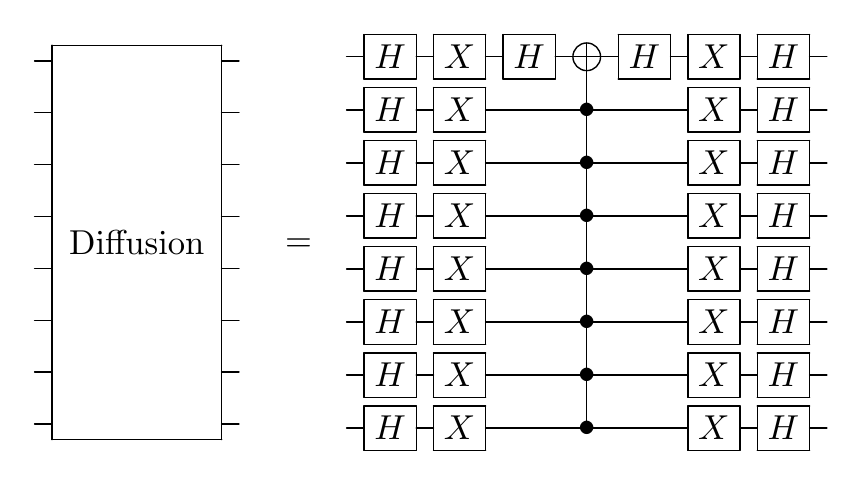}
    \caption{Explicit circuit construction that computes the inversion about the average on the quantum registers that encodes the superposition of all possible messages. This amplifies the amplitude of the correct answers.}
    \label{fig:diffusor}
\end{figure}

The explicit construction of the diffusion operator is common to all Grover implementations \cite{search-grover1996}. The role of this operator is to perform the inversion about the average once the states that codify the solutions of the problem have had the sign of their amplitude changed. The quantum circuit that achieves this is shown in \cref{fig:diffusor}. In this case, the diffusion operator only needs to be applied to the message registers of the quantum circuit as they are the only ones that are started in a superposition, in the proposed \toysponge this accounts for 8 qubits.
The diffusion operator corresponds to a matrix $D$ whose elements are
\begin{equation}
    D_{ij}=\frac{2}{N} \text{ if } i\neq j\quad\text{and}\quad D_{ii}=-1+\frac{2}{N},
\label{eq:diffusor}
\end{equation}
where $N=2^n$.

%Brief explanation of decomposing multi controlled CNOT gates into Toffoli gates
Both the oracle and the diffusion operator contain multi-CNOT gates that need to be decomposed into elementary gates in order to faithfully asses the full complexity of the circuit. Different methods in which multi-controlled gates can be decomposed in terms of CNOT and Toffoli gates are outlined and given their basic gate scaling in \cite{mCNOT-barenco1995}. Some of the most efficient constructions can only be performed in the case of having a circuit with some extra work qubits.

\begin{figure}
\centering
\begin{subfigure}[t]{0.36\textwidth}
    \includegraphics[width=\textwidth]{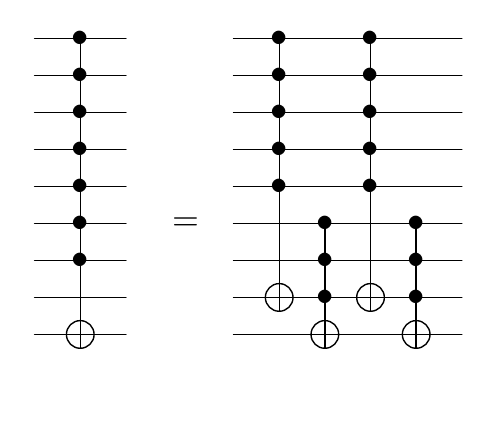}
    \caption{Multi-CNOT gate using an extra qubit}
    \label{fig:n-2_CNOT}
\end{subfigure}
\hfill
\begin{subfigure}[t]{0.63\textwidth}
    \includegraphics[width=\textwidth]{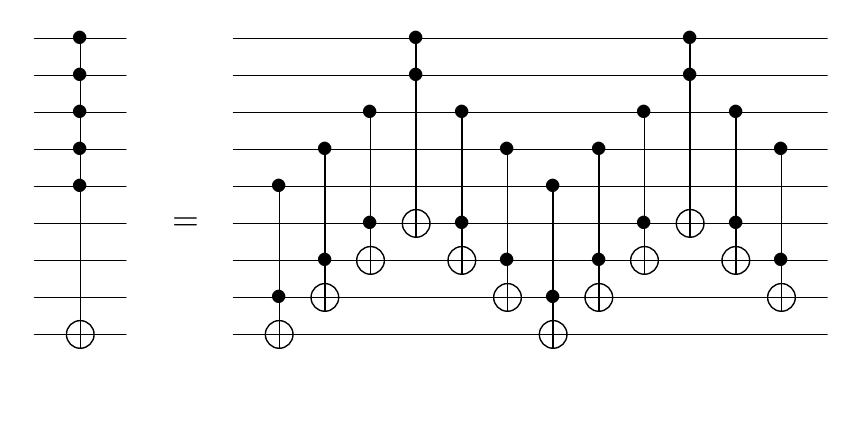}
    \caption{Multi-CNOT gate using extra work space.}
    \label{fig:n-m_CNOT}
\end{subfigure}
% \begin{figure}
% \centering
% \begin{subfigure}[b]{0.48\textwidth}
%     \input{figures/n-2_CNOT}
%     \caption{Multi-CNOT gate using an extra qubit}
%     \label{fig:n-2_CNOT}
% \end{subfigure}
% \begin{subfigure}[b]{0.48\textwidth}
%     \input{figures/n-m_CNOT}
%     \caption{Multi-CNOT gate using extra work space.}
%     \label{fig:n-m_CNOT}
% \end{subfigure}
\caption{Decomposition of multi-CNOT gates into basic Toffoli gates using a single extra work qubit. The extra qubit needs not be initialized at $\ket{0}$. Note that ancillas already required for the addition can be reused here. In Figure (a), the decomposition of the multi-CNOT gate with one work qubit into smaller gates is shown. In Figure (b), the full decomposition of the resulting gates is shown using enough work space so that they can be reduced to Toffoli gates.}
\label{fig:multiCNOT}
\end{figure}

% PACKAGE SUBFIGURE
% \begin{figure}
% \centering
% \resizebox{0.32\textwidth}{!}{
% \subfigure[\hspace{0.045cm} multi-CNOT gate using an extra qubit \label{fig:n-2_CNOT}]{
% \input{figures/n-2_CNOT}
% }
% }
% \hspace{0.05\textwidth}
% \resizebox{0.55\textwidth}{!}{
% \subfigure[\hspace{0.045cm} multi-CNOT gate using extra work space 
% \label{fig:n-m_CNOT}]{
% \input{figures/n-m_CNOT}
% }
% }
% \caption{Decomposition of multi-CNOT gates into basic Toffoli gates using a single extra work qubit. The extra qubit needs not be initialized at $\ket{0}$. Note that ancillas already required for the addition can be reused here. In Figure (a), the decomposition of the multi-CNOT gate with one work qubit into smaller gates is shown. In Figure (b), the full decomposition of the resulting gates is shown using enough work space so that they can be reduced to Toffoli gates.}
% \label{fig:multiCNOT}
% \end{figure}

A multi-CNOT gate can be decomposed, see \cref{fig:multiCNOT}, with linear efficiency into Toffoli gates using one extra qubit. There are in fact several unused qubits in the circuit when the multi-CNOT gates have to be applied, but in order to keep it separate from the qubits encoding the solutions, we shall use the ancillary qubit introduced in the addition modulo $2^n$ circuit as the work qubit for these construction.

%Since the adder modulo $2^n$ needs an ancillary qubit, we can perform this multi-CNOT gate that requires a single work qubit without needing to reuse one of the qubits that encodes parts of the hash construction. As \cref{fig:multiCNOT} shows, with a single extra work qubit, one can decompose a multi-CNOT gate using exclusively Toffoli gates.

\subsubsection{Full Grover step.}

\begin{figure}
\centering
\resizebox{0.5\textwidth}{!}{
\input{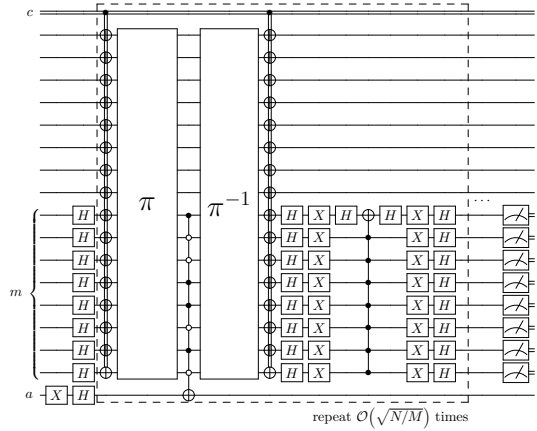}}
\caption{Explicit circuit construction that performs Grover's search algorithm in order to find preimages for a certain known Hash function following the \toysponge model described in \cref{sec:toy_sponge}. The label $c$ denotes an auxiliary classical register, $a$ is the Grover ancilla, and $m$ labels the qubit register that encodes the message. The dashed section of the circuit has to be repeated $\order{\sqrt{N/M}}$ times, where $N=2^n$ is the message space and M is the number of preimages, in order to complete the algorithm.}
\label{fig:grover_full}
\end{figure}

The Sponge Oracle and the diffusion operator combined to produce the body of a single Grover step and its full construction can be seen in \cref{fig:grover_full}. The full quantum circuit will require a series of $\order{\sqrt{N/M}}$ Grover steps, where $N= 2^n$ is the search space, and $M$ is the number of solutions, that is, preimages with the same hash value.  

With the full Grover step constructed, preimages of \toysponge can be obtained. This can be done directly if the number of preimages is known, applying the Grover step $\sim\frac{\pi}{4}\sqrt{2^n/M}$ times, with $M$ the total number of preimages. If that is not the case, this construction can be first employed in quantum counting algorithms in order to obtain the total number of preimages, following the guidelines illustrated in \cite{boyer1998tight}. 

\subsection{Unknown number of preimages.}

Alternatively, as shown in \cite{boyer1998tight} as well, this Grover step can be employed in an iterative algorithm to find a preimage even with an unknown number of solutions in the same order of complexity, that is $\order{\sqrt{2^n/M}}$ oracle calls. The algorithm assumes that the number of possible solutions is less than $3N/4$, where $N$ is the message space, which holds for hash functions as the number of preimages is small by construction.

The algorithm presented is as follows. First we initialize $m=1$ and $\lambda=6/5$, as described in \cite{boyer1998tight} any $\lambda$ between $1$ and $4/3$ would work. Then, a value $j$ is chosen randomly between the non-negative integers smaller than $m$ and Grover's search algorithm is applied with $j$ steps. If the measured outcome is a solution of the problem, the algorithm ends. If that is not the case, we set $m=\min(\lambda m, \sqrt{N})$ and repeat the previous step.

The success of this algorithm is not guaranteed in a set amount of steps unlike the regular Grover procedure. Nevertheless, the original paper proves that when the number of solutions is much lower than the total space, the number of Grover iterations is upper-bounded by $\frac{9}{4}\sqrt{N/M}$, where M is the number of solutions. The prefactor is larger, but even in the worst case, the overall scaling is still the same. However, the average number of function calls needed to solve the algorithm is much closer to the optimal scaling. In Sec. \ref{sec:results} we present average values of the iterations needed according to simulation.

\subsection{Scaling.}

%Scaling of a Grover step with this oracle and counting for the toy model and the real permutation
The explicit implementation of a toy example of a preimage searching algorithm using Grover's strategy can be used to extrapolate the scaling of said algorithm for real implementations.
Let us first state that the quantum circuit is made of many subparts that may be
implemented in some more efficient way in the future. In spite of that, it will not modify the complexity class of the problem.

%Following our implementation of a single Grover step, the number of TOFFOLI, CNOT and SINGLE qubit gates required as a function of the number of qubits $n$ is
By taking into account the scaling of each individual component of the circuit one can infer the full scaling of the circuit with regards to the size of the hash function used. The number of TOFFOLI, CNOT and SINGLE qubit gates required to build a Grover step for the \toysponge model are
\begin{equation}
    \text{\#TOFFOLI: }88 n-80s-88,\text{  \#CNOT: }240 n-160s,\text{  \#SINGLE: }84 n-160s+2,
\label{eq:gates}
\end{equation}
where $n$ is the total number of bits involved and $s$ is the number of sites in the permutation matrix. This scaling agrees with the construction of the full circuit in the code provided. Hash-specific alterations might add or subtract constants to these numbers but not in a significant way. The required depth of the algorithm turns out to be
\begin{equation}
    \text{depth} = \left(\frac{120}{\sqrt{s}}+8\right) n -120\sqrt{s}-80.
\label{eq:depth}
\end{equation}
This scaling benefits from the reduced depth of the addition and the fact that components of the circuit that act on different qubit registers can be applied at the same time.

Using the scaling we find for our \toysponge, it is possible to infer the amount of gates the algorithm would require to solve the same problem using the real permutation. In \cref{tab:scaling} the scaling of both the TOY model and REAL implementation is outlined.
\begin{table}
    \centering
    {\footnotesize
    \begin{tabular}{|c|c|c|c|c|c|c|}
        \cline{3-7}
         \multicolumn{2}{c|}{} & \#TOFFOLI & \#CNOT & \#SINGLE & \#TOTAL & depth \\
        \hline
        TOY &$n=16$ \,$s=4$ & 1000 & 3200 & 706 & 4906 & 1248 \\
        \hline
        REAL& $n=512$\, $s=16$ & 43688 & 120320 & 40450 & 204458 & 19856 \\
        \hline
    \end{tabular}
    }
    % \vskip .3cm
    \caption{Scaling of a single Grover step for the TOY and REAL model of preimage search based on the Sponge Hash using ChaCha20 permutation, as programmed in the algorithm presented above. The depth of the circuit is significantly lower than the total number of gates due to the possibility of applying quantum gates to different qubits in parallel. For the REAL implementation of the algorithm the depth reduction is increased when compared to the total gates since the larger permutation matrix allows for further parallel application of adder gates. This, however, would require a different ancillary qubit for each parallel adder circuit.}
    \label{tab:scaling}
\end{table}
This can be checked against the code provided \cite{hash-code} 
%\footnote{The reference to the GitHub repository is currently removed for anonymity.}.
for the example using real life values for the Sponge Hash construction.

This scaling of the basic Grover step has to be understood as the cost attached to the $\sqrt{2^n/M}$ oracle calls that the algorithms requires. The efficient construction of the oracle with the number of qubits signals that the overall scaling will not be altered, but it is a factor that has to be considered.

Some depth reduction can be achieved by implementing parts of the circuit that act on different sites of the permutation matrix at the same time. This has the cost of adding $\sqrt{s}$ ancillas for each adder modulo $2^n$ that can be implemented simultaneously. The extra qubits needed, however, are not significant when compared to the overall qubit complexity. In order to achieve the target circuit depth the number of qubits needed for both the TOY and REAL hash application are
\begin{equation}
    n_\text{TOY} = 19\text{ qubits},\quad n_\text{REAL} = 517\text{ qubits}.
\label{eq:qubits}
\end{equation}
The ancillary qubits needed for full depth reduction only account for less than a $1\%$ increase in the total size of the circuit for the real implementation.

\section{Quantum \toyblake construction}
\label{sec:attack_toy_blake}

The preimage finding algorithm for the more complex \toyblake construction, described in \cref{sec:toy_blake2} follows the same basic outlines as the ones described for the \toysponge model. In this case, since the message is used inside the \toyblake compression function, it has to be kept in a separate register than the permutation matrix $v$. This causes the amount of qubits needed to increase when compared to \toysponge. In spite of that, Grover's search algorithm follows as previously described, with the diffusion operator acting exclusively on the message space.

\begin{figure}[ht]
    \centering
    \resizebox{0.75\textwidth}{!}{
    \input{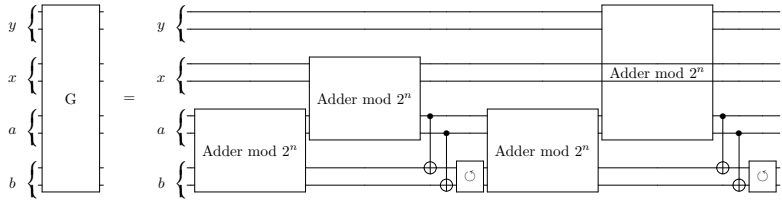}
    }
    \caption{Quantum circuit analog of operator $G$ presented in \cref{sec:toy_blake2}. Each presented wire corresponds to two quantum wires in the toy construction, and any CNOT gate between wires is understood to carry onto omitted wires. The adder mod $2^n$ gates are equivalent to the ones presented in \cref{fig:adder}. In all instances, the register $a$ is the one updated with the addition result.}
    \label{fig:quantum-g}
\end{figure}

The internal mixing function for the \toyblake implementation as a reversible quantum circuit is showcased in \cref{fig:quantum-g}. The basic parts that make up this function are the same as for the $\chachaperm$ Quarter Round, but here there is the contribution from registers $x$ and $y$, which are part of the message. 

\begin{figure}
    \centering
    \resizebox{0.75\textwidth}{!}{
    \input{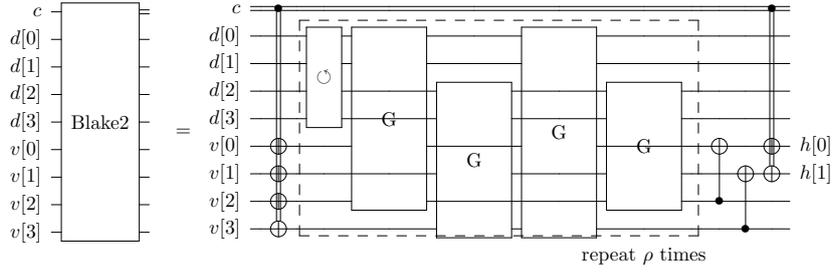}
    }
    \caption{Quantum circuit that implements \toyblake compression function as is outlined in \cref{fig:blake2_toy2x2_compression_function}. The initialization step is performed classically and stored in a classical register $c$ along with the value of $iv$ as they are used as classical controls for the quantum circuit. The shuffling in the message space translates to a qubit relabelling and does not incur any quantum gates. Each visible quantum wire in the figure accounts for 4 qubits in the toy implementation. The hash value is returned in the $v[0]$ and $v[1]$ registers.}
    \label{fig:quantum-blake-comp}
\end{figure}

The compression function that is seen in \cref{fig:blake2_compression_function} can be built as a reversible quantum circuit for Grover's oracle using as many qubits as bits in the message matrix $d$ and vector $v$. A single ancillary qubit, or two if further parallelization is desired, is needed for the addition modulo $2^n$ circuits and it can then be reused throughout the repetition of the $G$ function operators. The first step of vector $v$ can be initialized classically, as it is fully determined by the initial parameters of the system, and stored in a classical register to act as classical controls later on in the system. The permutation of the message matrix can be achieved through a qubit relabeling and thus does not add more gates. In \cref{fig:quantum-blake-comp} a sketch of the circuit is shown. In the figure, each quantum wire represents a quantum register of 4 different qubits and the gates are understood to act equally on all of them.

\begin{figure}[ht]
    \centering
    \resizebox{0.5\textwidth}{!}{
    \input{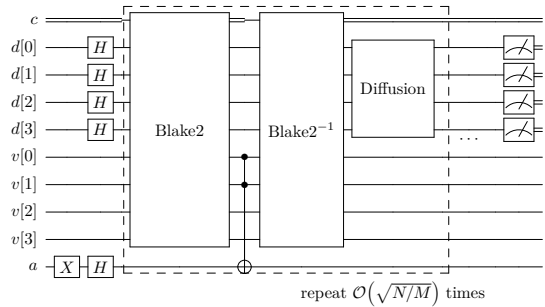}}
    \caption{Grover's algorithm construction for preimage finding for \toyblake presented in \cref{sec:toy_blake2}. Each quantum wire labeled $d$ and $v$ correspond to 4 qubits matching the description in their respective section. Labelled $c$ is the classical register used in the Blake2 operator. A gate acting on the quantum wire in the figure is to be understood as acting on all qubits that it represents. The hash value is encoded into the controls that target the Grover ancilla from registers $v[0]$ and $v[1]$. The Grover step has to be repeated $\order{\sqrt{N/M}}$ times where $N=2^n$.}
    \label{fig:blake-grover}
\end{figure}

Once the compression function is built, an oracle can be constructed to find preimages of a specific hash for this toy construction. The explicit circuit of the Grover implementation can be seen in \cref{fig:blake-grover}, with quantum wires labeled $d$ and $v$ representing quantum registers of 4 qubits. Accounting for the extra registers, the construction is identical to the one showcased for the \toysponge model in the previous section. The diffusion operator is constructed in the same way as depicted in \cref{fig:diffusor}, accounting for circuit size differences so as to match an operator with the elements described in \cref{eq:diffusor}.
The amount of qubits needed to perform the toy model of this algorithm is $32$ qubits for the message and vector registers, as well as the Grover ancilla and the required ancilla for the modular additions. Thus, the minimum number of qubits needed to run the algorithm is 34.

\subsection{Scaling.}

An explicit gate-by-gate implementation of the Grover iteration for preimage finding for \toyblake has been constructed. Due to the overall size of the circuit, efficient classical simulation is impractical. However, from the construction we have presented, the complexity scaling can be calculated and therefore extrapolated to real implementations. 

In terms of basic gates, the Grover step that has to be repeated in order to find preimages for \toyblake requires the following basic gates,
\begin{align}
    \text{\#TOFFOLI: }&(8\rho+16\frac{\rho}{\sqrt{s}}+12)n-(8s+16\sqrt{s})\rho-56,\nonumber\\
    \text{\#CNOT: }&(24\rho+40\frac{\rho}{\sqrt{s}}+1)n-(16s+32\sqrt{s})\rho,\\
    \text{\#SINGLE: }&(8\rho+16\frac{\rho}{\sqrt{s}}+7)n-(16s+32\sqrt{s})\rho+2,\nonumber
    \label{eq:blake2-scaling}
\end{align}
where $n$ is the amount of bits of the message digest $d$, $s$ is the total sites of the permutation matrix, and the parameter $\rho$, as seen in \cref{fig:blake2_compression_function}, determines the amount of repetitions inside the compression function. Hash-specific alterations might add or subtract constants to these numbers, but not relevant to the overall scaling. The circuit can achieve a certain level of parallelization provided $\sqrt{s}$ ancillas are added to the circuit. Then, the depth of the circuit becomes 
\begin{equation}
    \text{depth}=(12\frac{\rho}{\sqrt{s}}+16\frac{\rho}{s}+16)n-(12\sqrt{s}+24)\rho-50.\\
\end{equation}
Further detailed in \cref{tab:blake2-scaling} are the total amount of gates required for the TOY implementation described beforehand, and for a REAL instance of $1024$ bits, with 64 bit words in the sites of the matrix. For both circuits, the amount of repetitions of the \toyblake Quarter rounds has been set to $\rho = 12$. 
\begin{table}
    \centering
    {\footnotesize
    \begin{tabular}{|c|c|c|c|c|c|c|}
       \cline{3-7}
         \multicolumn{2}{c|}{}& \#TOFFOLI & \#CNOT & \#SINGLE & \#TOTAL & depth \\
        \hline
        TOY &$n=16$ \,$s=4$ & 2240 & 6928 & 1650 & 11018 & 1550 \\
        \hline
        REAL& $n=1024$\, $s=16$ & 157384 & 414208 & 150018 & 721610 & 64622 \\
        \hline
    \end{tabular}
    }
    % \vskip .3cm
    \caption{Scaling of a single Grover step for the TOY and REAL model of preimage search based on the BLAKE2. Both computations have been done with the same $\rho=12$. To achieve the presented depth, $\sqrt{s}$ ancillas are needed.}
    \label{tab:blake2-scaling}
\end{table}
Again, this scaling refers to the Grover step, oracle and diffuser included, and can be understood as the factor that accompanies each necessary application of the Grover step in the algorithm. 
This can be corroborated with the code provided \cite{hash-code}
%\footnote{The reference to the GitHub repository is currently removed for anonymity.}.
for the example using real life values for the Blake2 Hash construction.

The qubit complexity of the \toyblake implementation, as briefly discussed above, is higher than for the \toysponge model based on the $\chachaperm$ permutation. This is due to the fact that the messages are actively used inside the \toyblake Quarter Rounds, therefore are to be saved on their own quantum register, effectively doubling the number of qubits required. The total amount of qubit width required to run the presented algorithm, including the $\sqrt{s}$ ancillas in order to achieve full parallelization are
\begin{equation}
    n_\text{TOY} = 35\text{ qubits},\quad n_\text{REAL} = 2053\text{ qubits}.
\label{eq:blake2-qubits}
\end{equation}
The toy model is at the limit of what can be implemented by quantum simulators available, 
but is within the number of qubits available in recent quantum devices.

\section{Results}
\label{sec:results}

The explicit construction and programming of all steps in a Grover attack on \toysponge allows for a detailed exact simulation of the results and costs of an attack on a hash function. In the following we will discuss the success in the finding of preimages, according to the expected performance of the algorithm in ideal conditions, that is, without
introducing a simulation of the experimental errors which are expected.
Furthermore, the fact that we handle the exact description of the state at every step of the computation opens the possibility to analyze the entanglement entropy which is pervading the system. This is, in turn, makes it possible to assess the limits of an approximate simulation of a quantum circuit using Tensor Networks.
A study of the effects of Pauli errors, appearance of random $X$, $Y$ or Z gates after any gate application, is performed in order to compare the effectiveness of running the full algorithm or a reduced version under noise conditions.

\subsection{Finding preimages.}

%Graphs of the probabilities of finding solutions to a hash for different hash lengths
Let us first run the attack on \toysponge using the quantum circuit we have constructed in the previous section. The algorithm is made of a sequence of basic Grover steps. According to the theory, even if the exact number of preimages is unknown at first, it is necessary to apply $\order{\sqrt{N/M}}$ Grover steps \cite{boyer1998tight}, where $N=2^n$ and $M$ is the number of preimages, in order to find those preimages.
Every Grover step will increase the probability of the desired solution, until the maximum is obtained. 

\begin{figure}
    \centering
    \begin{subfigure}[b]{0.49\textwidth}
        \includegraphics[width=\textwidth]{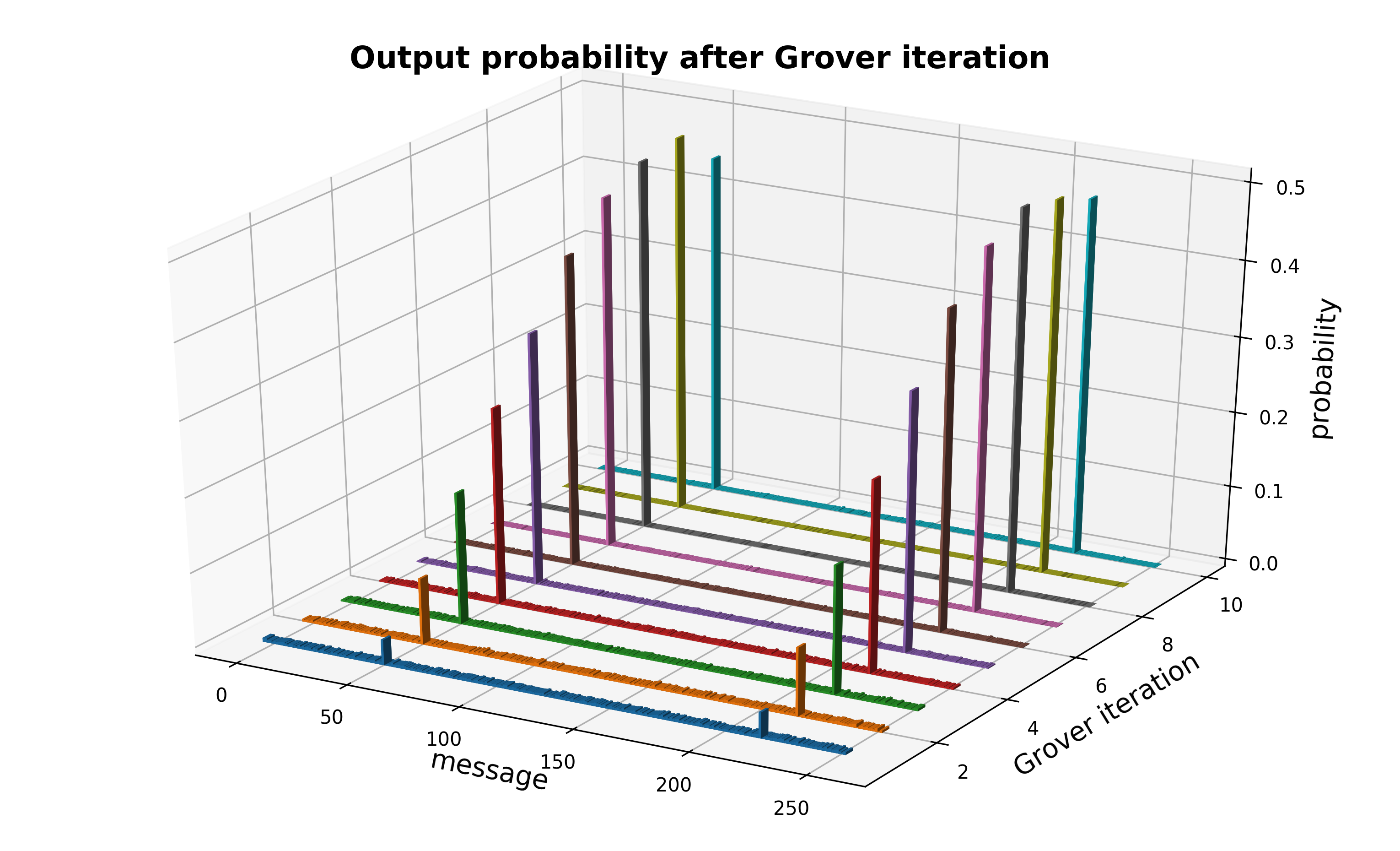}
        \caption{Instance with two preimages.}
        \label{fig:grover_2preim}
    \end{subfigure}
    \hfill
    \begin{subfigure}[b]{0.49\textwidth}
        \includegraphics[width=\textwidth]{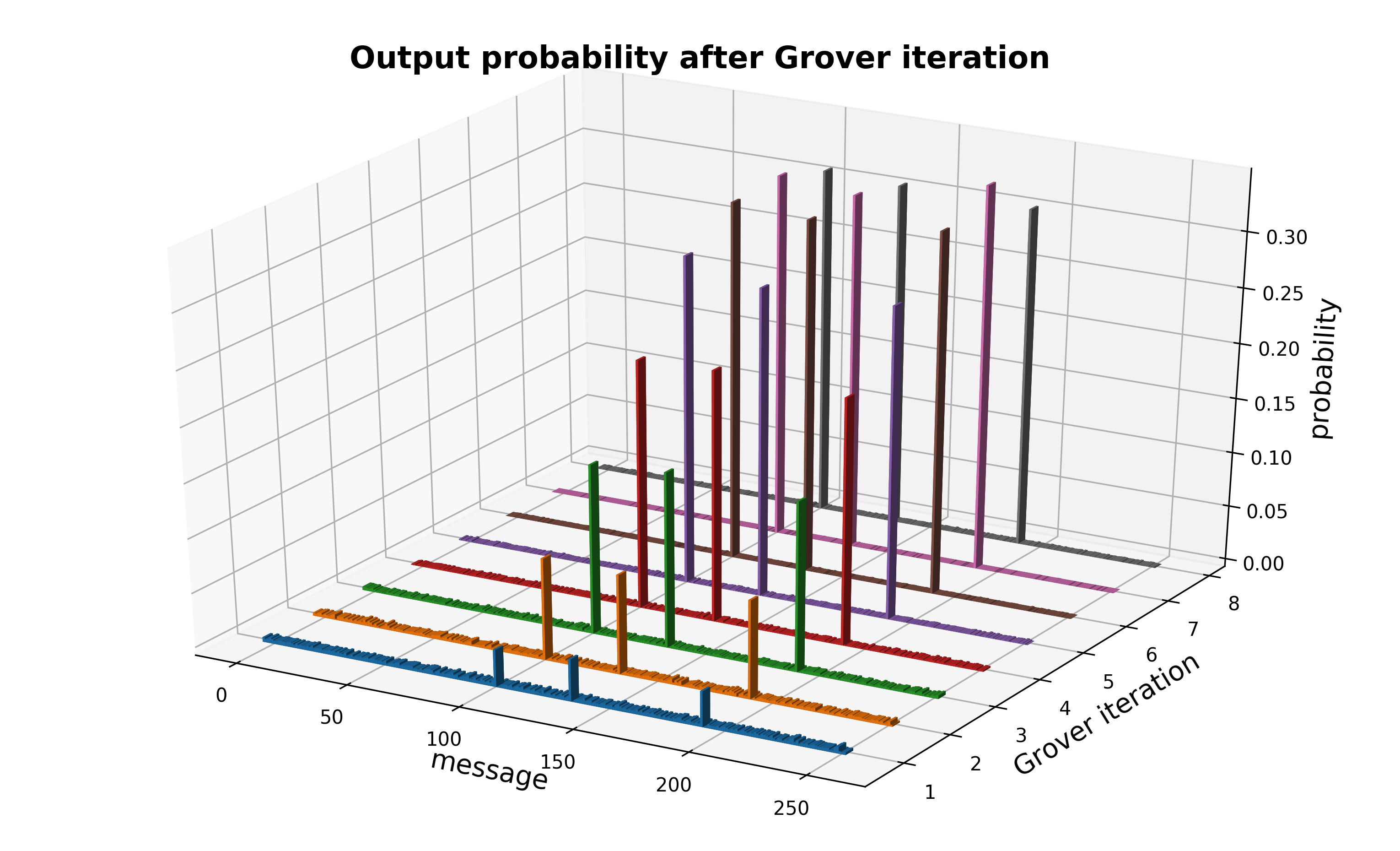}
        \caption{Instance with three preimages.}
        \label{fig:grover_3preim}
    \end{subfigure}
    \caption{Evolution of the probability of finding a message during the Grover process. As more Grover steps are performed the probability of finding a preimage of the target hash is amplified. As can be appreciated in both images, after just a portion of the required $\sim\frac{\pi}{4}\sqrt{2^n/M}$ Grover steps, the solutions become easily noticeable.}
    \label{fig:grover_evolution}
\end{figure}

In order to visualize the iterative nature of Grover's algorithm, we plot in \cref{fig:grover_evolution} the probabilities of measuring the final message states for hash instances with two and three preimages respectively, as a function of Grover steps. It can be seen that the probability of measuring the preimages is amplified with each iteration following a sinus wave pattern reaching its maximum at the closest integer near $\sim\frac{\pi}{4}\sqrt{2^n/M}$, as expected. After that point, the probability of finding the solutions decreases as it is redistributed back to all states. Grover's algorithm can be understood as a rotation in the two dimensional plane defined by a vector with the superposition of all solutions and another orthogonal vector with the superposition of the non-solution states. 

% PACKAGE SUBFIGURE
% \begin{figure}
%     \centering
%     \subfigure[\hspace{0.005cm} instance with two preimages \label{fig:grover_2preim}]{\includegraphics[width=0.49\textwidth]{figures/grover_it_10100011.png}}
%     \subfigure[\hspace{0.005cm} instance with three preimages \label{fig:grover_3preim}]{\includegraphics[width=0.49\textwidth]{figures/grover_it_1010100.png}}
%     \caption{Evolution of the probability of finding a message during the Grover process. As more Grover steps are performed the probability of finding a preimage of the target hash is amplified. As can be appreciated in both images, after just a portion of the required $\sim\frac{\pi}{4}\sqrt{2^n/M}$ Grover steps, the solutions become easily noticeable.}
%     \label{fig:grover_evolution}
% \end{figure}

%Discussion on the number of Grover steps performed vs. the samples extracted from the system
As it can be appreciated in \cref{fig:grover_evolution}, there is no need to reach the full number of Grover steps in order to already find an important amplification of the probability of measuring a preimage state. This implies that one could stop the quantum computation before the full scaling is reached, and extract more output samples in a way that a solution will still be found with high probability. Outlined in \cref{tab:prob_apparition} are the probabilities of a preimage appearing after each amount of Grover steps, as well as the average number of samples needed to find the first preimage apparition. This strategy to cut short the full quantum computation has been analyzed for several number of preimages. It can be seen that the optimal way to proceed in an ideal quantum computer is to finish the full Grover's algorithm and perform all iterations. However, if unlimited circuit depth is not available, as is the case for Noisy Intermediate-Scale Quantum (NISQ) \cite{preskill2018quantum} devices where gate errors and decoherence are a relevant issue, one can strive for a set depth and still arrive to the right solution by extracting more samples.
This practical consideration may be non-trivial in this and other applications of Grover's search algorithm.
\begin{table}[ht]
    \centering
    {\footnotesize
    \begin{tabular}{|p{1.2cm}|p{0.8cm}|p{0.9cm}|p{0.8cm}|p{0.8cm}|p{0.9cm}|p{0.8cm}|p{0.8cm}|p{0.9cm}|p{0.8cm}|}
        \cline{2-10}
        \multicolumn{1}{c|}{} & \multicolumn{9}{c|}{Number of preimages} \\
        \cline{2-10}
        \multicolumn{1}{c|}{} & \multicolumn{3}{c|}{2} & \multicolumn{3}{c|}{4} & \multicolumn{3}{c|}{6} \\ 
        \cline{2-10}
         \multicolumn{1}{c|}{}& Prob. & First & Calls & Prob. & First & Calls & Prob. & First & Calls \\ 
         \hline
         1 step & 0.069 & 14.523 & 15 & 0.135 & 7.417 & 8 & 0.198 & 5.052 & 6 \\ 
         \hline
         2 steps & 0.183 & 5.453 & 11 & 0.344 & 2.908 & 6 & 0.483 & 2.067 & 5 \\ 
         \hline
         3 steps & 0.337 & 2.966 & 9 & 0.591 & 1.691 & 6 & 0.774 & 1.291 & 4 \\ 
         \hline
         4 steps & 0.511 & 1.956 & 8 & 0.816 & 1.225 & 5 & 0.965 & 1.036 & 5 \\ 
         \hline
         5 steps & 0.684 & 1.463 & 8 & 0.964 & 1.038 & 6 & 0.986 & 1.015 & 5 \\ 
         \hline
         6 steps & 0.834 & 1.120 & 8 & 0.997 & 1.003 & 6 &  & & \\
         \hline
         7 steps & 0.942 & 1.062 & 8 &  &  &  &  & & \\ 
         \hline
         8 steps & 0.996 & 1.004 & 8 &  &  &  &   & &\\ 
         \hline
    \end{tabular}
    }
    % \vspace{0.3cm}
    \caption{Probability of success and number of average samples needed before finding one preimage depending on the number of Grover steps performed. The average number of total oracle calls in order to find a preimage is also presented. For different number of preimages, the probability of measuring one of them increases according to the number of Grover steps applied. In general, the optimal solution is to perform the full Grover's algorithm to find solutions. However, should circuit depth be an issue, we can stop at an earlier step and acquire more samples for a similar result.}
    \label{tab:prob_apparition}
\end{table}

\subsection{Function calls needed for unknown number of solutions.}

The implementation of Grover's algorithm when the exact number of solutions is unknown does not have a strict number of function calls needed to find a preimage. The algorithm presented in \cite{boyer1998tight} gives a strict upper-bound on the number of Grover steps needed. In the following we have applied this algorithm to the \toysponge construction presented here in order to estimate the average amount of Oracle calls needed for instances with different number of preimages.

\begin{table}[ht]
    \centering
    {\footnotesize
    \begin{tabular}{|p{3cm}|p{1.5cm}|p{1.5cm}|p{1.5cm}|}
        \cline{2-4}
        \multicolumn{1}{c|}{} & \multicolumn{3}{c|}{Number of preimages} \\
        \cline{2-4}
        \multicolumn{1}{c|}{} & \multicolumn{1}{c|}{2} & \multicolumn{1}{c|}{4} & \multicolumn{1}{c|}{6} \\
        \cline{2-4}
         \hline
         Grover scaling & 8 & 6 & 5  \\ 
         \hline
         Average calls & 10.0 & 6.3 & 4.7  \\ 
         \hline
         Upper-bound & 26 & 18 & 15  \\
         \hline
    \end{tabular}
    }
    % \vspace{0.3cm}
    \caption{Theoretical bounds for the number of function calls needed to recover a solution in the \toysponge construction compared with the average calls needed in a simulation. The average is taken after performing the algorithm $1000$ times. The first value presented is the scaling of the direct Grover algorithm, the second one is the averaged value from simulation, and the third one is the theoretical upper bound described in \cite{boyer1998tight}. It can be seen that the average number of oracle calls needed to solve the problem is much closer to the normal Grover scaling than to the predicted upper bound.}
    \label{tab:unknown-m}
\end{table}

In Tab. \ref{tab:unknown-m} we present data on the average number of Oracle calls needed to recover one preimage when the total number of solutions is not known. The average calls is obtained after averaging over 1000 different iterations of the algorithm. As seen, the average number of steps remains much closer to the original Grover scaling rather than the theoretical upper bound predicted.

\subsection{Entropy obstruction to simulation by Tensor Networks.}

A study of the entropy within the quantum circuit is relevant, as classical techniques are available to simulate quantum algorithms in a most efficient manner if the entanglement present along a quantum computation is low. As a matter of fact, the technology usually quoted as Tensor Networks is known to achieve a faithful representation of any quantum state with moderate entanglement. 

Let us introduce the von Neumann entropy as a figure of merit to quantify entanglement in the register $\ket{\psi}$. The way to compute the entropy of a bi-partition $A-B$ of the system requires to first 
get the reduced density matrix to half of the register $\rho_A= {\rm Tr}_B |\psi\rangle\langle\psi|$. Then the von Neumann entropy for this partition is
% \begin{equation}
    $S_A=-{\rm Tr} \rho_A \log_2 \rho_A \ .$
% \end{equation}
This entropy is zero in the absence of entanglement and is bounded by the size of the system $n_A$, that is the number of qubits in the partition. 
It is known that
states whose von Neumann entropy only scales logarithmically with the number of qubits can be described in terms of Matrix Product States\cite{efficientmps}. This means the amount of entropy present in the quantum register along the quantum circuit should be large, otherwise there would exist an efficient attack on hash functions based on simulating Grover's algorithm with Tensor Networks.

It is well known that the entropy present in the register along Grover's algorithm is bounded by 1 if measured after an application of each Grover step \cite{entanglement-orus2004}. This has a very simple explanation. At the end of a Grover step, the register is separated in two distinct orthogonal states, the solutions and the rest. As a consequence, the maximum von Neumann entropy between circuit bi-partitions reached its bounded by $1$, corresponding to maximum two-partite entropy. But this argument fails to understand that the heart of the quantum computation is done within the oracle. There, the quantum register displays an enormous increase of entropy, which is at the origin of its quantum advantage.

Let us emphasize this point further. No algorithm that does not produce large entropy can provide any quantum advantage over classical strategies. This is due to the fact that a low amount of entanglement can be simulated efficiently. Grover's advantage needs to be rooted at exploiting large entanglement in the quantum register. This, indeed, should take place within the oracle. 

\begin{figure}
    \centering
    \begin{subfigure}[b]{0.4\textwidth}
        \includegraphics[width=\textwidth]{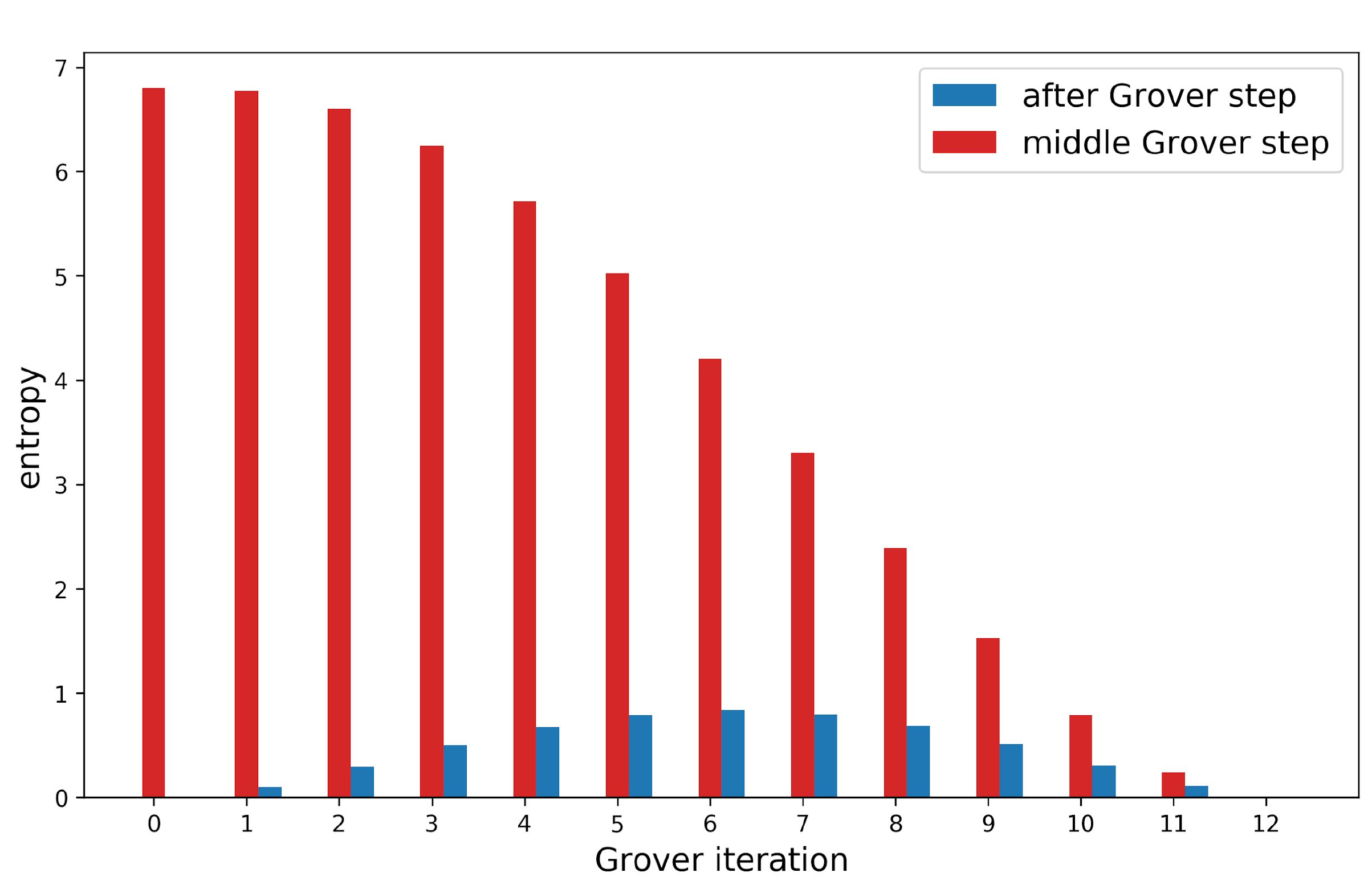}
        \caption{Hash: 10011101 - 1 preimage.}
        \label{fig:ent_1preimage}
    \end{subfigure}
    % \hfill
    \begin{subfigure}[b]{0.4\textwidth}
        \includegraphics[width=\textwidth]{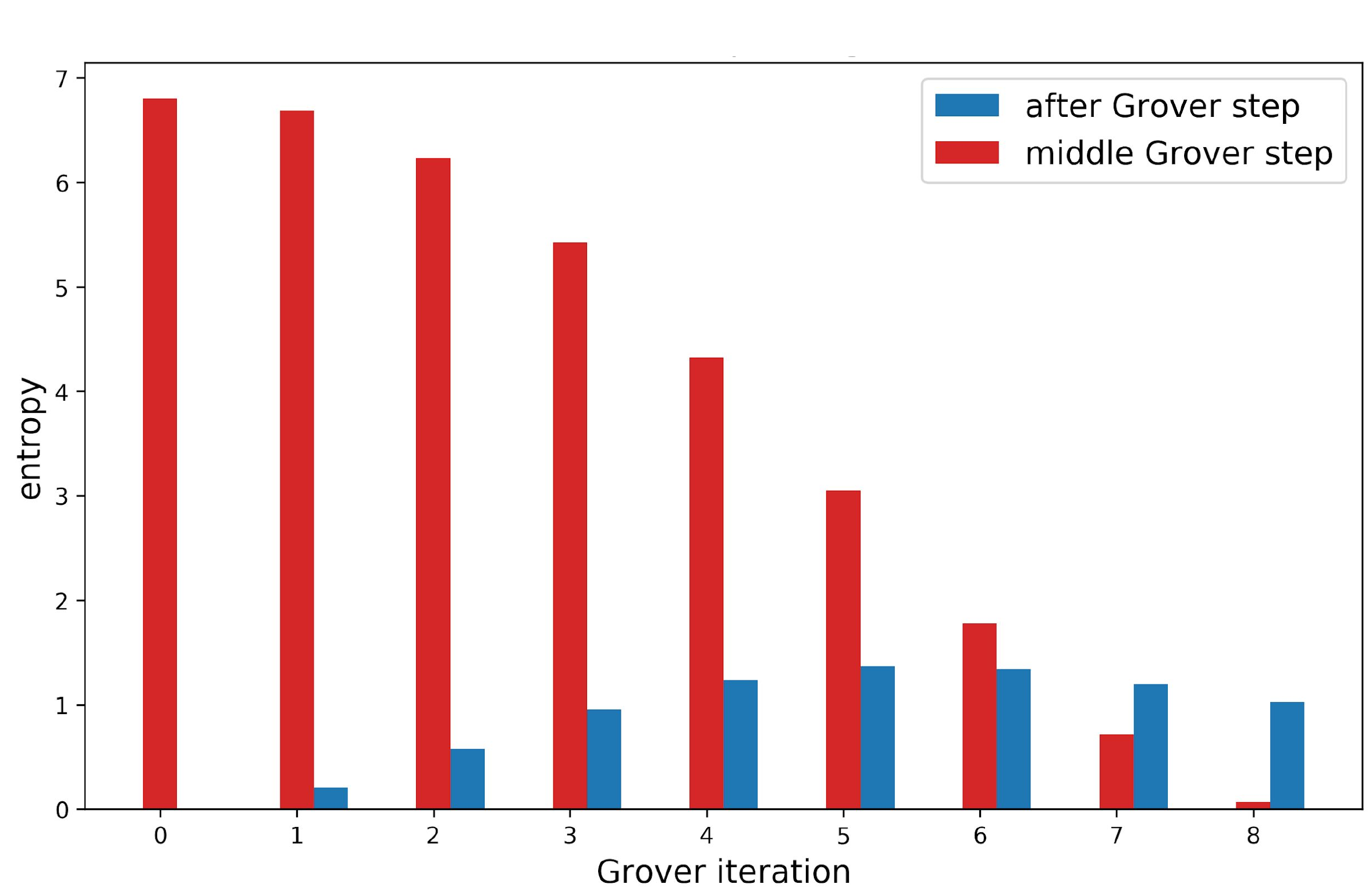}
        \caption{Hash: 10100011 - 2 preimages.}
        \label{fig:ent_2preimage}
    \end{subfigure}
    % \hfill
    \begin{subfigure}[b]{0.4\textwidth}
        \includegraphics[width=\textwidth]{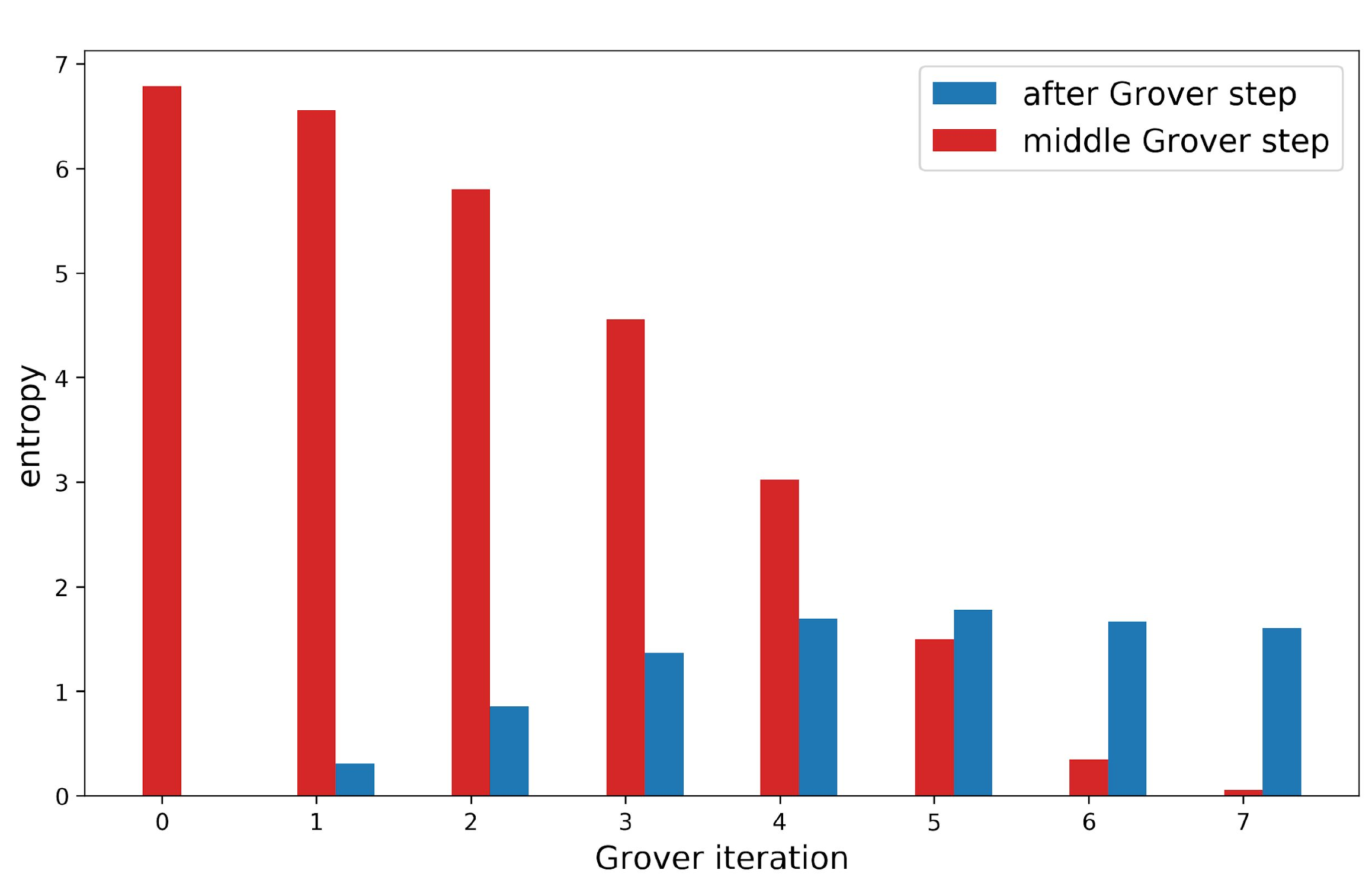}
        \caption{Hash: 10000101 - 3 preimages.}
        \label{fig:ent_3preimage}
    \end{subfigure}
    % \hfill
    \begin{subfigure}[b]{0.4\textwidth}
        \includegraphics[width=\textwidth]{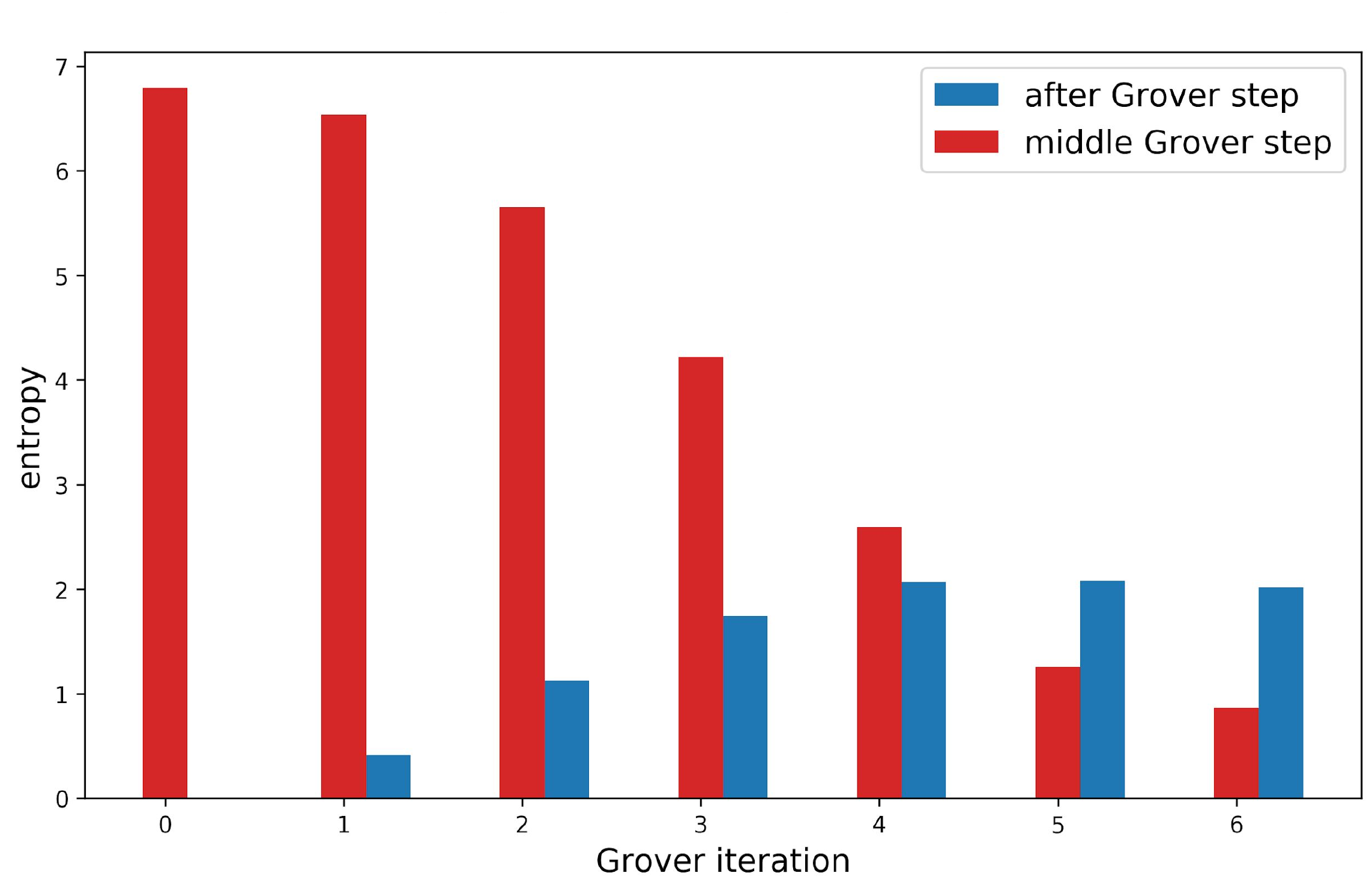}
        \caption{Hash: 1110001 - 4 preimages.}
        \label{fig:ent_4preimage}
    \end{subfigure}
    \caption{Von Neumann entropy in the middle and after a Grover step for different number of preimages. In this analysis, all  bipartitions remain orthogonal in the final solution state. In the case where entropy is measured after Grover step we consider the bipartition that corresponds to half the message space. In the case where entropy is measured in the middle of the Grover step, we consider the bi-partition of half the qubits that correspond to the whole permutation matrix.}
    \label{fig:entropy}
\end{figure}

A confirmation of this reasoning can be seen in \cref{fig:entropy}. The von Neumann entropy for half of the register in the middle of the Grover step is closer to the maximum possible bipartite entropy in the \toysponge. More precisely, the entropy in the middle of the Grover step is computed right after the Grover ancilla changes the sign of the target hash states, at the center of the circuit shown in \cref{fig:oracle}. The bi-partition used at this point is half the quantum registers that make up the permutation matrix. For the entropy after the Grover step, the bi-partition is instead half the quantum registers that encode the message space, as the rest of the permutation matrix is at the $\ket{0}$ state.

% PACKAGE SUBFIGURE
% \begin{figure}
%     \centering
%     \subfigure[\hspace{0.045cm} hash: 10011101 - 1 preimage \label{fig:ent_1preimage}]{\includegraphics[width=0.35\textwidth]{figures/entropy-1.pdf}}
%     \subfigure[\hspace{0.045cm} hash: 10100011 - 2 preimages \label{fig:ent_2preimage}]{\includegraphics[width=0.35\textwidth]{figures/entropy-2.pdf}}
%     \subfigure[\hspace{0.045cm} hash: 10000101 - 3 preimages \label{fig:ent_3preimage}]{\includegraphics[width=0.35\textwidth]{figures/entropy-3.pdf}}
%     \subfigure[\hspace{0.045cm} hash: 1110001 - 4 preimages \label{fig:ent_4preimage}]{\includegraphics[width=0.35\textwidth]{figures/entropy-4.pdf}}
%     %
%     \caption{Von Neumann entropy in the middle and after a Grover step for different number of preimages. In this analysis, all  bipartitions remain orthogonal in the final solution state. In the case where entropy is measured after Grover step  we consider the bipartition that corresponds to half the message space. In the case where entropy is measured in the middle of the Grover step, we consider the bi-partition of half the qubits that correspond to the whole permutation matrix.}
%     \label{fig:entropy}
% \end{figure}

In the \toysponge quantum algorithm we have presented, a gate by gate study of the entanglement entropy reveals that its maximum is 7.3619, regardless of the number of preimages. This is to be compared with the theoretical maximum, 8. It is noteworthy to observe that the maximum along the computation is 
reached at the first action of the Sponge Oracle. 
In some way, the register develops very large correlations which are needed to spot the solutions. Then, as the solutions are enhanced, quantum correlations need not be that high.

The entanglement in the register along the computation depends on the number of preimages as well as if their binary encoding is orthogonal in the respective bi-partitions. Nevertheless, entanglement always peaks at the first application of the Sponge Oracle. This shows that, in the case of large number of qubits $n$, the entropy in the register will likely scale with $n$, rendering inefficient the classical alternatives for simulation of quantum circuits. Grover's algorithm does need an actual quantum computer to support entropy that scales as the volume of the system.

\subsection{Error robustness.}

The previous simulations assume access to an error free quantum device. This, however, is not the case for actual quantum computers, still far from operating with logical qubits. Before reaching fault-tolerant quantum computation quantum devices will be in their NISQ stage, so considering the effects of errors in proposed algorithms is required. In the following we analyse the effect of errors on these implementations of Grover's algorithm for preimage finding. 

Quantum errors range from decoherence and dephasing to Pauli errors. Pauli errors are characterized by the probability of a random Pauli effect, represented as gates X, Y and Z, acting on any particular qubit. We have simulated the \toysponge construction in the presence of increasing probability of appearing an X, Y and Z gate after every quantum gate is applied.

\begin{figure}[ht]
    \centering
    \includegraphics[width=0.75\textwidth]{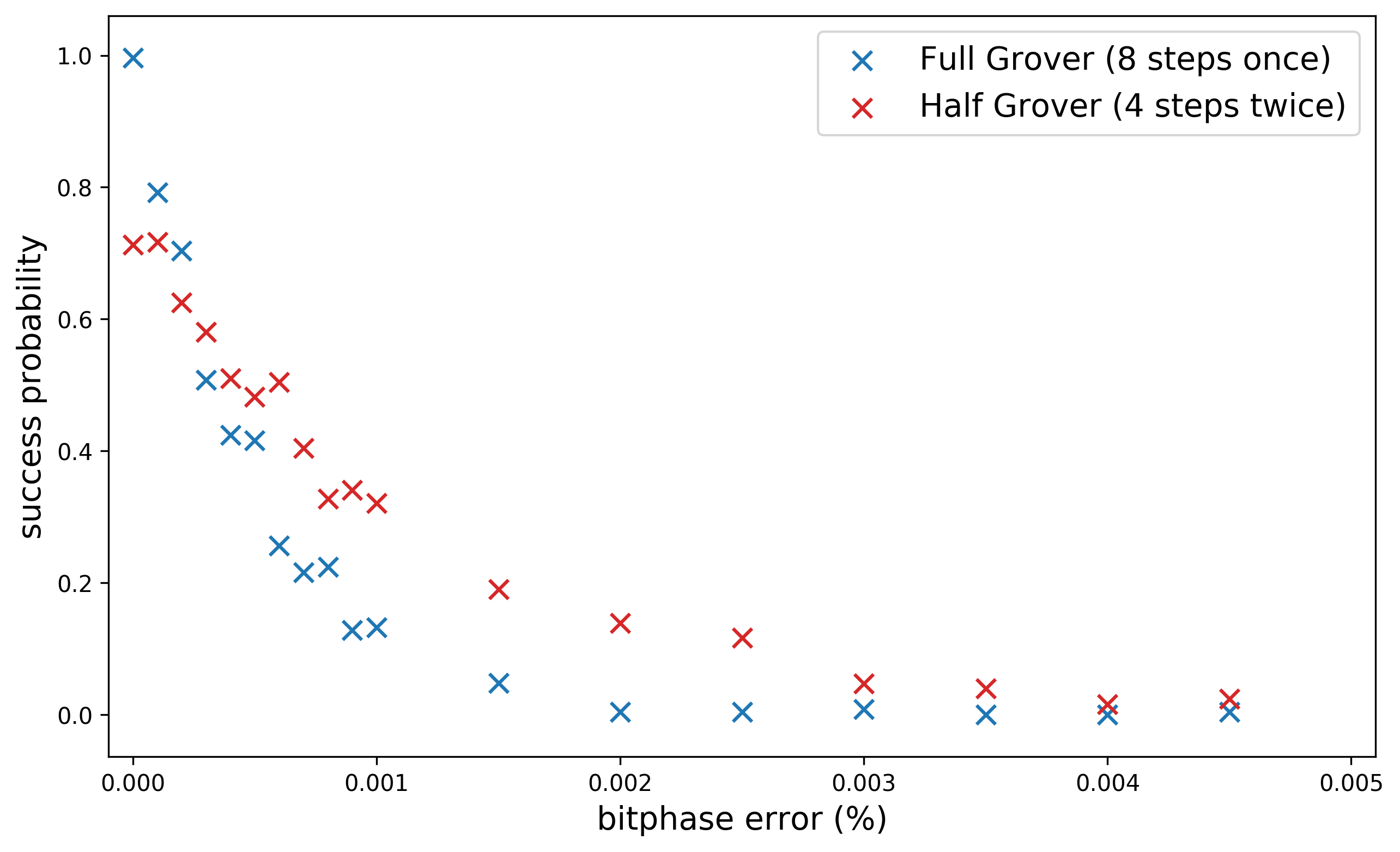}
    \caption{Probability of finding a preimage after applying Grover's algorithm with the same amount of oracle calls under increasing bitphase, or Pauli, error. A small error already destroys most of the strength of the algorithm due to its complexity and required precision. Points averaged over 250 instances. Running multiple instances of a reduced version of the algorithm can result in mitigation of the error effects on Noisy Intermediate-Scale Quantum computers. Examples run for an instance with two preimages corresponding to hash value: 10100011.}
    \label{fig:bitphase-err}
\end{figure}
Due to the amount of required gates for Grover's algorithm, this toy model requires gate precision below $0.001\%$ in order to reliably output the correct preimages. When the Pauli error probability reaches beyond $0.003\%$ all results are effectively lost. In \cref{fig:bitphase-err} the probability of recovering a preimage after applying Grover's algorithm for different magnitudes of Pauli errors is shown. For the same amount of oracle queries, results are presented for the full Grover implementation and one where the preimages are measured when half the Grover steps are applied. It can be appreciated that for no error, applying all steps, namely $\sim\frac{\pi}{4}\sqrt{2^n/M}$, gives better results. However, as the errors increase, the reduced gate number of applying less Grover steps translates into an advantage when finding preimages. For this example, the success probability doubles when at $0.001\%$ error probability. This balance should be taken into consideration when running this algorithm on a real device. As the Hash construction is scaled, at least a similar ratio of error probability to number of required gates should be maintained, and since the number of gates needed is much larger, the error has to go down that much as well. This leads to the conclusion that even though performing less Grover steps yields a marginal advantage against error, it is negligible when full size Hash constructions are considered.

\section{Conclusion}
\label{sec:conclusions}

Quantum algorithms specially designed to attack cryptographic protocols require extensive perusal. We have presented an explicit quantum code that performs a full Grover attack on two scaled ARX-based hash functions. Our algorithm is simulated using the Qibo quantum simulation software, which is an open-source library for quantum computation available in \cite{qibo20}. Code to reproduce the quantum circuits present in this paper can be found in GitHub \cite{hash-code}.
%will be uploaded to GitHub\footnote{The reference to the GitHub repository is currently removed for anonymity.}.
%Examples of the circuits developed in this paper can be found in GitHub \cite{CircuitExamples}, in OpenQASM form, so they can be run in a multitude of available quantum simulation software.

The explicit construction of a quantum algorithm based on Grover's search provides an exact quantification of the quantum resources which are needed for an actual implementation of a preimage attack. The main takeaway is that the number of gates and depth required to build the oracle and diffusor scales linearly with $n$, the number of qubits in the register, with large prefactors. This is relevant as large prefactors may make inaccurate naive predictions for the power of quantum computation. 

The results of this work further bring a number of considerations we shall now summarize: (a) Different hash functions may differ substantially in the quantum resources needed to attack them. The quantum algorithm to find preimages of a BLAKE2 hash requires 2053 qubits and 721610 gates for a single Grover step, while for the basic sponge construction using ChaCha20 permutation 517 qubit and 204458 gates are required (considering a single compression function/permutation cycle). (b) Classical gates such as AND, OR or XOR are dealt differently at the quantum level. It turns out that XOR easily translates onto a CNOT. But, a classical AND or OR gate is not reversible which implies proliferation of qubits. Thus, new hash functions can be designed to be more difficult to be attacked by a quantum Grover strategy. (c) The analysis of the entropy that the register develops shows that entanglement is maximal during the first action of the oracle. This fact discards the possibility of simulating the quantum algorithm using the powerful Tensor Network classical techniques. (d) We have shown that the idea of sampling intermediate Grover steps is proven essentially as powerful as running the whole algorithm in an ideal quantum computer. (e) We have presented a first analysis of Pauli errors, showing the degradation of the success probability as the probability of error increases. We have also observed that a strategy to run part of Grover's algorithm performs better than the full quantum circuit because of the smaller accumulation of errors.

\bibliographystyle{alpha}
\bibliography{Citations}

\end{document}